\documentclass[twocolumn,tighten,twocolappendix]{aastex631}
\hypersetup{colorlinks,linkcolor=blue,citecolor=blue,urlcolor=blue}
\usepackage{bm}
\usepackage{graphicx}
\usepackage{epsf}
\usepackage{graphics}
\usepackage{amsmath}
\usepackage{subfigure}
\usepackage{enumitem}
\usepackage{physics}
\usepackage{color, xcolor}

\usepackage{soul} 
\soulregister{\cite}7 
\soulregister{\citep}7
\soulregister{\citet}7 
\soulregister{\ref}7 
\soulregister{\pageref}7

\def \revised{ }

\def \Pi{$t_{\rm 12}$}

\def\aap{Astrophysical Journal}

\def\apjs{ApJS}

\newcommand\dex{\,{\rm dex}}
\def\beq{\begin{equation}}
\def\eeq{\end{equation}}
\def\bey{\begin{eqnarray}}
\def\eey{\end{eqnarray}}
\newcommand\Kelvin{\,{\rm K}}

\def\kms{\,{\rm {km\, s^{-1}}}}
\def\Msun{\,{\rm M_\odot}}

\newcommand\perccm{\, {\rm cm}^{-3}}

\newcommand\yr{\,{\rm yr}}
\def\Myr{\,{\rm Myr}}
\def\Gyr{\,{\rm Gyr}}

\def\gs{\mathrel{\raise1.16pt\hbox{$>$}\kern-7.0pt
\lower3.06pt\hbox{{$\scriptstyle \sim$}}}}
\def\ls{\mathrel{\raise1.16pt\hbox{$<$}\kern-7.0pt
\lower3.06pt\hbox{{$\scriptstyle \sim$}}}}
\def\gtsima{\, {\buildrel > \over \sim} \,}
\def\ltsima{\, {\buildrel < \over \sim} \,}
\def\prosima{\, {\buildrel \propto \over \sim} \,}
\def\gsim{\lower.5ex\hbox{\gtsima}}
\def\lsim{\lower.5ex\hbox{\ltsima}}
\def\simgt{\lower.5ex\hbox{\gtsima}}
\def\simlt{\lower.5ex\hbox{\ltsima}}
\def\simpr{\lower.5ex\hbox{\prosima}}

\shorttitle{Halo-Galaxy-SMBH Co-Evolution}
\shortauthors{H. Li et al.}

\begin{document}

\title{Physical Processes Behind the Co-Evolution of Halos, Galaxies and Supermassive Black Holes in the IllustrisTNG Simulation}

\correspondingauthor{Hao Li, Yangyao Chen, Huiyuan Wang}
\email{lh123@mail.ustc.edu.cn, yangyaochen.astro@foxmail.com, whywang@ustc.edu.cn}

\author[0000-0002-4326-3543]{Hao Li}
\affiliation{
Department of Astronomy, 
University of Science and Technology of China, 
Hefei, Anhui 230026, People's Republic of China;}
\affiliation{
School of Astronomy and Space Science \\
University of Science and Technology of China, Hefei 230026, China}

\author[0000-0002-4597-5798]{Yangyao Chen}
\affiliation{
Department of Astronomy, 
University of Science and Technology of China, 
Hefei, Anhui 230026, People's Republic of China;}
\affiliation{
School of Astronomy and Space Science \\
University of Science and Technology of China, Hefei 230026, China}

\author[0000-0002-4911-6990]{Huiyuan Wang}
\affiliation{
Department of Astronomy, 
University of Science and Technology of China, 
Hefei, Anhui 230026, People's Republic of China;}
\affiliation{
School of Astronomy and Space Science \\
University of Science and Technology of China, Hefei 230026, China}

\author[0000-0001-5356-2419]{Houjun Mo}
\affiliation{
Department of Astronomy\\
University of Massachusetts, Amherst MA 01003, USA}


\begin{abstract}
We explore the co-evolution of dark matter halos, their central galaxies, and 
central supermassive black holes (SMBHs) using the IllustrisTNG (TNG) simulation. 
We find that the evolutionary histories of individual galaxies in the 
$M_{\rm BH}$-$M_*$ plane can be decomposed into four distinct 
phases, separated by three transition points. 
We identify the driving processes of galaxy evolution within each phase and 
derive the conditions necessary and sufficient for transitions to subsequent phases. 
The first phase is dominated by star formation, with its duration primarily 
determined by the mass of the SMBH seed and the surrounding gas environment. 
The second phase is characterized by rapid SMBH growth, 
and the transition to the next phase occurs when the thermal-mode feedback of 
active galactic nucleus (AGN) can unbind gas from the galaxy. 
The third phase involves self-regulation of the SMBH, and the transition 
to the quenched phase occurs when the kinetic-mode feedback of 
AGN counterbalances gas cooling within the subhalo. 
The final phase is dominated by mergers. 
We investigate the use of scaling relations among different 
mass components and evolutionary phases to understand 
processes implemented in TNG and other simulations, and discuss
how current and forthcoming observations can be used to constrain models.
\end{abstract}


\section{Introduction}\label{sec_intro}

Structure formation in the $\Lambda$CDM cosmology is predicted to follow a 
hierarchical paradigm \citep[e.g.][]{peeblesLargescaleStructureUniverse1980,
blumenthalFormation1984,moGalaxyFormationEvolution2010}, 
in which structures grow from smaller to larger scales, forming virialized 
dark matter halos of larger masses. Gas associated with dark matter, 
which may be heated by shocks associated with gravitational collapse, 
can cool radiatively and contract until it becomes self-gravitating and 
fragments to form stars. Under extreme conditions that are most likely
fulfilled in the central regions of galaxies, the gas and/or stars can  
be turned into black holes (BHs); the latter can continue to grow 
via accretion of gas and/or mergers with stars and other BHs. This may lead 
to the formation of supermassive black holes (SMBHs) observed in the centers of many 
galaxies \citep[e.g.][]{
Kormendy2013,
grahamAppreciatingMergersUnderstanding2023,
maiolinoSmallVigorousBlack2024}. 
In this scenario of structure formation, small and compact objects in the cosmic 
hierarchy are embedded in the environments defined by the more extended and less dense 
structures. Inversely, energetic processes on small scales can affect their 
environments through feedback processes, forming a bidirectional connection in the hierarchy.
Examples of such feedback include the AGN feedback from an accreting SMBH,
which can push/heat the gas within its host galaxy 
\citep[e.g.][]{fabianObservationalEvidenceActive2012,
liuObservationsFeedbackRadioquiet2013,heckmanCoevolutionGalaxiesSupermassive2014,
kuboAGNIonizedGas2022,
donahue2022BaryonCyclesBiggest} 
and halo \citep[e.g.][]{liELUCID2022,luoELUCIDVIIISimulating2024,liu2024XrayCoolCore},
and the feedback from star formation, which can change the composition and 
state of the interstellar and circumgalactic medium 
\citep[e.g.][]{fieldingImpactStarFormation2017,
hopkinsRadiativeStellarFeedback2020,liEffectsSubgridModels2020,
dengSimulatingIonizationFeedback2024} and even modify the distribution
of dark matter \citep[e.g.][]{ElZant_etal2001,MoMao2004,  
agertzIMPACTLARFEEDBACK2016,
acevedoDarkMatterInducedBaryonic2023,yangDarkMatterMeasurements2024}. 
The formation and evolution of SMBHs, galaxies and halos are thus 
indispensable to one another, and have to be modeled together in order to
understand the physical processes behind observational phenomena.

Numerous theoretical efforts have been made to model structure formation
in the current cosmological framework. 
Numerical \citep[e.g.][]{
    schayeEAGLEProjectSimulating2015,
    hopkinsFIRE2SimulationsPhysics2018,
    daveSimba2019a,
    pillepichFirst2018,
    blankNIHAOXXIIIntroducing2019},
semi-analytical \citep[e.g.][]{
    whiteCoreCondensationHeavy1978a, 
    somervilleSemianalyticModelCoevolution2008a, 
    henriquesGalaxyFormationPlanck2015, crotonSEMIANALYTICGALAXYEVOLUTION2016,
    baughGalaxyFormationPlanck2019},
analytical \citep[e.g.][]{
    moFormationGalacticDiscs1998,
    hernquistAnalyticalModelHistory2003, 
    sharmaIkaModelFeedbackregulated2020,
    bowerDark2017b, 
    dekelEfficientFormationMassive2023}
and empirical \citep[e.g.][]{
    zhengTheoreticalModelsHalo2005,
    hearinIntroducingDecoratedHODs2016,
    xuConditionalColourMagnitude2018,
    moTwophaseModelGalaxy2024,chenTwophaseModelGalaxy2024,chenTwophasePaper3-2024}
models have been 
developed to describe the physical 
processes on different scales and to discover, test and verify the assumptions 
and prescriptions governing these processes.  Various observations
of SMBHs, galaxies and halos have been used to constrain these models. 
Examples of observations include the spatially resolved properties of individual
galaxies and clusters \citep[e.g.][]{
linSDSSIVMaNGAInsideout2019,
fahrionFornax3DProject2020,
jingOriginQuenchedGasrich2024}; 
the spatial number density of galaxies
binned by different quantities, such as the galaxy luminosity or stellar mass 
function, and AGN luminosity or SMBH mass function 
\citep[e.g.][]{
chenELUCIDVICosmic2019,
padmanabhanConstrainingAGNLuminosity2023,
wangTrueNumberDensity2024}; the spatial 
correlation of galaxies and AGNs \citep[e.g.][]{liDependenceClusteringGalaxy2006,
garcia-berneteGalaxyActivityTorus2024,saxenaWidespreadAGNFeedback2024}; the joint distribution or 
scaling relations between different quantities, such as the SMBH
mass-stellar mass relation \citep[e.g.][]{Kormendy2013,
grahamAppreciatingMergersUnderstanding2023,Hong2023}, 
the SMBH mass-halo mass relation \citep[e.g.][]{
zhang2024HaloMassobservableProxy,li2024BlackHoleHalo}, and 
the stellar mass-halo mass relation 
\citep[e.g.][]{Yang2007,Yang2008,Moster2010,Wechsler2018,
Zhang2022}.

Comparisons between observations and models have led to many important
conclusions about the physical processes on different scales and the interaction
between them. For example, the star formation efficiency, defined as the 
ratio of stellar mass to halo mass, is found observationally
to be an unimodal function of halo mass, peaked at around a halo mass of 
$M_{\rm h,c} \approx 10^{12} M_{\odot}$ 
\citep[e.g.][]{Yang2003,yang2012EVOLUTIONGALAXYDARK,behrooziAVERAGESTARFORMATION2013}.
The peak star formation efficiency is about a few percent, much lower 
than the value of $\approx 0.16$ allowed by the cosmic baryon fraction.
Two different channels of feedback, one from 
stars \citep[e.g.][]{Kauffmann1998,Hopkins2012,Hopkins2014,
angles-alcazarBlackHolesFIRE2017,hopkinsWhyBlackHoles2021a} 
and the other from AGNs \citep[e.g.][]{bowerDark2017b,weinbergerSupermassiveBlackHoles2018,daveSimba2019a} are thus 
proposed to suppress the star formation, and are inferred to be dominant
at low- and high-masses, respectively.
Another example is the relation between SMBH mass and galaxy bulge/total 
stellar mass, which is observed to roughly follow a power 
law \citep[e.g.][]{Kormendy2013,
grahamAppreciatingMergersUnderstanding2023, grahamReading2023, Hong2023}.
Different mechanisms, such as star formation and SMBH growth regulated by 
AGN feedback \citep[e.g.][]{DiMatteo2005,weinbergerSupermassiveBlackHoles2018,Hong2023}, 
and the galaxy-galaxy and SMBH-SMBH mergers \citep[e.g.][]{Peng2007a},
have been proposed to explain the relation, 
in particular the power-law index. Precise measurements of the relation over 
a wide range of redshift from observations are thus critical to distinguish 
theoretical models, although such measurements are difficult to make at high redshifts 
\citep[e.g.][]{Schaye2010, Kormendy2013, greeneIntermediateMass2020}.
The stellar mass-halo mass relation and the SMBH mass-stellar mass relation
also imply a correlation in mass between SMBHs and halos,
as indeed found observationally by using halo masses obtained from the abundance 
matching \citep{behrooziUniverseMachineCorrelationGalaxy2019}, satellite 
kinematics \citep{ferrareseBulgeFundamentalRelation2002}, 
gravitational lensing \citep{zhang2024HaloMassobservableProxy,li2024BlackHoleHalo}, 
and luminosity in X-ray \citep[e.g.][]{gaspariXRayHaloScaling2019,voitBlackHoleGrowth2024},
and SMBH mass from velocity dispersion \citep[e.g.][]{Zahid2016,Zahid2018,Sohn2020,
Shankar2020}.
{\revised Although the correlation between SMBH mass and halo mass may be just a 
phenomenological coincidence, physical frameworks involving the synergy
between halo potential and AGN feedback in regulating the baryon cycle
from SMBH scales to halo scales have been proposed to interpret the 
correlation \citep[e.g.][]{bowerDark2017b,voitBlackHoleGrowth2024},
and observational evidences supporting the interpretation 
have been identified 
\citep[e.g.][]{bandaraRELATIONSHIPSUPERMASSIVEBLACK2009,
bogdanCONNECTINGDARKMATTER2015,zhang2024HaloMassobservableProxy}.
}

James Webb Space Telescope (JWST) has opened a new era to study
galaxy formation at high redshift. Over the past two years, 
observations with JWST have discovered a large number of galaxies that are much 
different from those seen in the local Universe. 
{\revised Galaxies at high redshift appear to be more
irregular and clumpy \citep[e.g.][]{maSimulatingGalaxiesReionization2018,kuhnJWSTRevealsSurprisingly2024,
hiranoEarlyStructureFormation2024,bikClumpyStarFormation2024},
with some dominated by dense star-forming clumps 
\citep[e.g.][]{mowlaFireflySparkleEarliest2024, adamoDiscoveryBoundStar2024}.
As the coupling of feedback energy from accreting SMBHs is expected to largely 
depend on the morphology \citep{sivasankaranAGNFeedbackIsolated2025} and 
clumpiness \citep{duttaDissipationAGNJets2024} of the host galaxy,
SMBHs at high redshift may follow the relations with their host galaxies
and halos that are different from those in the local Universe.}
Meanwhile, an abundant population of ``little red dots'' (LRDs),
featured by reddened spectra and small sizes,  has been discovered at $z \gtrsim 4$, 
and has been inferred as AGNs and/or young stars obscured by dust
\citep{mattheeLittleRedDots2024,perez-gonzalezWhatNatureLittle2024,liLittleRedDots2024}.
{\revised If LRDs are indeed AGNs, this population of SMBHs is likely to reside 
in galaxies in a special stage of the evolution, and extends our 
observational picture of actively accreting SMBHs to the range of 
low BH mass in the early Universe.}
In addition, a recent analysis by \citet{wangStrongHeII2024} targeting at a young 
galaxy at $z\approx 8$ has suggested that its emission lines and spectral 
slope are most consistent with a cluster of Pop-III stars. The massive 
stars within this cluster are thus candidates for SMBH seeds.
The above observations with JWST qualitatively support the theoretical
expectations of galaxy formation in the $\Lambda$CDM paradigm, and, 
for the first time, join the pieces of information throughout the full growth 
histories of galaxies and SMBHs from their infancies to the present day.  
To use such observations to constrain theoretical models, it is crucial to 
identify processes that are responsible for salient observational 
properties, and to understand how different processes are linked together to 
produce the populations of galaxies and SMBHs observed at any given time.  

In this paper, we use the hydrodynamical simulation, IllustrisTNG, 
to study the co-evolution of SMBHs, galaxies and halos. We quantify the growth 
histories of individual systems, break each history into a series of distinct 
phases by a set of transition points, and identify the driving processes of each phase. 
The results found in this paper are specific to IllustrisTNG, but the method 
can be applied to other theoretical models as a unified way to  
compare models with each other and with observations.

This paper is organized as follows. In \S\ref{sec:data-and-sample}, we introduce
the simulation data and sample. In \S\ref{sec:def-phases}, we define the 
phases and transitions. In \S\ref{sec:processes}, we analyze the co-evolution
of SMBHs, galaxies and halos in each phase, identify the driving processes,
and derive the conditions for the transition to the next phase. 
In \S\ref{sec:observation}, we discuss the potential use of observations
to constrain the models. In \S\ref{sec:summary-and-discussion}, 
we summarize our results.

\section{Data and samples}
\label{sec:data-and-sample}    

In this paper, we use data from the IllustrisTNG project 
\citep{Marinacci2018,Naiman2018,nelsonFirst2018,pillepichFirst2018,springelFirst2018,nelsonIllustrisTNGSimulationsPublic2019} 
to investigate the co-evolution of dark matter halos, galaxies, 
and supermassive black holes (SMBHs). The physical and subgrid models adopted 
by IllustrisTNG are detailed in the method papers of the project
\citep{weinbergerSimulatingGalaxyFormation2017,pillepichSimulatingGalaxyFormation2018a} 
together with specific improvements over the original Illustris simulation
\citep{vogelsbergerModel2013,genelIntroducingIllustrisProject2014,
sijackiIllustrisSimulationEvolving2015}. 

To achieve a balance between sample size and numerical resolution, we 
use the IllustrisTNG100-1 run (hereafter TNG), which has a simulation box 
of $\approx 110.7$ comoving $\rm Mpc$ on each side. Galaxy formation in the box is 
simulated with $2\times 1820^3$ baryon elements, with a mass resolution of 
$m_{\rm baryon}=1.4\times 10^6 \Msun$.
TNG adopted a flat $\Lambda$CDM cosmology, with parameters taken from the Planck2015 
results \citep{planckcollaboration2016Planck2015Results}: 
$h=H_{0}/(100\ \rm km/s/Mpc)=0.6774$, $\Omega_{\rm M,\,0}=0.3089$, 
$\Omega_{\rm B,\,0}=0.0486$, and $\Omega_{\rm \Lambda,\,0}=0.6911$. 

Groups and galaxies are identified from the simulation, and information 
about individual halos and galaxies is provided. Dark matter halos are selected 
using the friends-of-friends (FoF) algorithm with a linking length $b=0.2$ 
(in units of the mean separation of dark matter particles).
Dark matter subhalos are identified using the {\sc Subfind} algorithm \citep[e.g.][]{springelPopulatingClusterGalaxies2001,rodriguez-gomez2015MergerRateGalaxies}
to connect all of the gravitationally bound particles or cells. 
Baryon particles are linked to the same group as their closest dark matter 
particles, and galaxies are defined as the baryon components 
of subhalos in FoF-groups. Typically albeit not always \citep[e.g.][]{nelsonIllustrisTNGSimulationsPublic2019}, the most massive 
subhalo within a halo is defined as the central subhalo, and the galaxy 
associated with it is referred to as the central galaxy. 
All subhalos other than the central within a halo 
are ``satellites''. Throughout this paper, 
we only consider subhalos of a cosmological origin defined by the {\tt SubhaloFlag}, 
as described in the data release of TNG \citep{nelsonIllustrisTNGSimulationsPublic2019}.

A baryonic version of the {\sc SubLink} algorithm is applied to TNG to build the 
subhalo merger trees \citep[see][for details]{rodriguez-gomez2015MergerRateGalaxies}.
Each subhalo is linked with a set of progenitor subhalos, among which the one 
with the ``most massive history'' is defined as the first/main
progenitor. 
To trace the evolution of a halo, its central galaxy, and the central SMBH 
within the central galaxy, we follow the main progenitor of the central subhalo 
recursively back in time. This extracts a branch of subhalos from the whole 
merger tree rooted in the subhalo in consideration, and we refer to it as the 
main branch of this root subhalo/galaxy/SMBH. 

TNG has provided a list of available properties for each halo, subhalo, galaxy 
and SMBH \citep[see][]{nelsonIllustrisTNGSimulationsPublic2019}. For other properties required
by this study but not provided in the data release, we compute them from the 
particles/gas cells. The properties used in the paper are listed as follows.
\begin{itemize}[topsep=0pt,parsep=0pt,itemsep=0pt,leftmargin=0pt]
    \item 
    $\bm M_{\rm h}$: halo mass, defined as the total mass enclosed in a sphere 
    centered on the particle of minimal gravitational potential 
    among all particles in the 
    FoF halo, with a radius, $R_{\rm h}$, defined so that the mean 
    density within the sphere is 200 times the mean density of the universe
    at the epoch when the halo is identified. The circular velocity of the 
    halo at $R_{\rm h}$ is defined as the virial velocity $V_{\rm h}$.
    \item
    $M_*$: stellar mass of the galaxy, defined as the sum of the masses of 
    the bound stellar particles enclosed in a sphere centered on the particle 
    with minimum gravitational potential among all particles within the subhalo, 
    with a radius equal to $2 R_{\rm *,1/2}$, twice of the stellar half-mass 
    radius of the galaxy.
    \item 
    $M_{\rm BH}$: {\revised the mass of the most massive black-hole particle 
    in the subhalo. }
    \item 
    $M_{\rm gas}(<R)$: the sum of masses of all bound gas cells enclosed in 
    a sphere centered on the same location as that for $M_*$, 
    with a radius $R$.
    \item 
    $V_{\rm max}$: the maximum of the spherically-averaged rotation curve,
    accounting for the gravity of all species of particles/cells. 
    \item 
    $\dot{M}(t_n)$: the growth rate of mass $M$ at the $n$-th snapshot, 
    estimated by central difference as
    \begin{equation}
        \dot{M}(t_n)= \frac{M(t_{n+1})-M(t_{n-1})}{t_{n+1}-t_{n-1}}\,,
    \end{equation}
    where $t_{n-1}$, $t_{n}$ and $t_{n+1}$ are the cosmic times of the 
    previous, current, and next snapshots.
    \item 
    $\dot{M}(t_n)/M(t_n)$: the specific growth rate of mass $M$ at 
    the $n$-th snapshot. 
\end{itemize}
For the halo, stellar and SMBH masses, the definition of $\dot{M}(t_n)$ yields 
the halo accretion rate (HAR), the star formation rate (SFR), and 
the black hole accretion rate (BHAR), respectively, and 
the definition of $\dot{M}(t_n)/M(t_n)$ yields their specific values:
the specific halo accretion rate (sHAR), the specific black hole accretion rate (sBHAR), 
and the specific star formation rate (sSFR). {\revised
Note that the integration of the growth rate defined here is ensured to be
equal to the mass whose evolution pattern is the main focus of this paper.}

{\revised To tell whether a galaxy is star-forming or quenched, we find its instantaneous
star formation rate by summing over all gas cells within $2R_{*,1/2}$, and obtain the 
instantaneous sSFR by dividing the instantaneous SFR by $M_*$.
We note that the instantaneous SFR is noisy due to the stochastic nature
of the subgrid models, the finite resolution of the simulation
and the defects of the structure finder algorithms that link the 
resolution elements.
Some galaxies, in particular the high-redshift ones, can thus have sudden drops 
in sSFR in some individual snapshots, make an excursion to the ``quenched'' state 
and return to the ``star-forming'' status later.
To mitigate the noise, we smooth the instantaneous sSFR along the main 
branch of the subhalo merger tree by a running Gaussian 
kernel with standard deviation equal to the dynamical timescale 
($t_{\rm dyn, h} \equiv R_{\rm h}/V_{\rm h}$) of the 
host halo at each snapshot.
We define a galaxy to be quiescent/quenched if the sSFR so obtained
is less than the threshold of $10^{-11}\,{\rm yr}^{-1}$, which is commonly used in literatures
\citep[e.g.][]{schayeEAGLEProjectSimulating2015,
cuiOriginGalaxyColour2021,habouzitSupermassiveBlackHoles2021,
bluck2024GalaxyQuenchingHigh}.
For each quenched galaxy, we identify its quenching time, $t_{\rm q}$, as 
the first time when its smoothed instantaneous sSFR
falls below the threshold of $10^{-11}\,\yr^{-1}$.
}


\begin{figure}
\centering
\includegraphics[width=.45\textwidth]{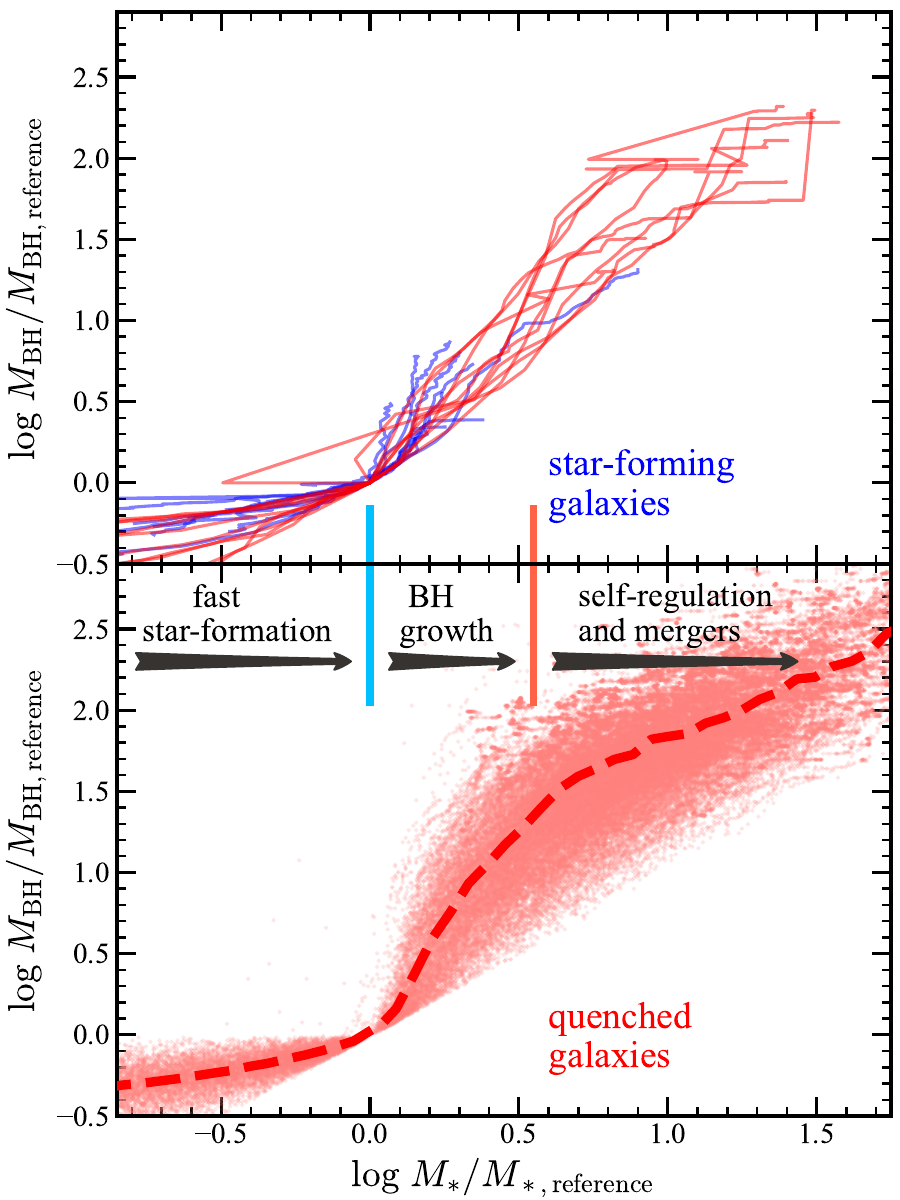}
\caption{The main-branch histories of individual galaxies in the 
    $M_{\rm BH}$-$M_*$ plane. For each branch, $M_*$ and $M_{\rm BH}$ are 
    scaled by their values at a reference redshift when the ratio,
    $M_{\rm BH}/M_*$, reaches its minimum.
    {\bf Top panel} shows the histories of randomly selected galaxies in TNG 
    at $z=0$ with $M_*>10^{8.5}\Msun$, 10 for star-forming ({\bf blue}) and 10 
    for quenched ({\bf red}).
    {\bf Bottom panel} shows the sample of quenched galaxies used in the 
    analysis throughout this paper. Here, each {\bf red dot} represents
    a galaxy at a snapshot in a main branch. The {\bf red dashed curve} shows
    the median among all branches. See \S\ref{sec:data-and-sample} for
    the details of sample and selection. 
}
\label{fig:sample-overall} 
\end{figure}

Fig.~\ref{fig:sample-overall} shows the main-branch histories of individual
galaxies in the $M_{\rm BH}$-$M_*$ plane. The top panel plots the histories  
of galaxies randomly selected from all central galaxies with $M_* > 10^{8.5}\Msun$ 
at $z=0$, 10 for star-forming (blue) and 10 for quenched (red).
Here, for each history, $M_*$ and $M_{\rm BH}$ are scaled by their values 
at a reference redshift when $M_{\rm BH}/M_*$ reaches its minimum.
For each history, the $M_{\rm BH}$-$M_*$ curve presents regular patterns 
that mark the different phases undergone by the galaxy. In the early time before 
the reference point, the curve is quite flat, indicating that the growth of SMBH is 
slower than that of the stellar component. A sharp transition appears at about the reference 
point, followed by a rapid growth of SMBH. In the later time, the difference between 
star-forming and quenched galaxies appears. For quenched galaxies, the growth of 
$M_{\rm BH}$ slows down, and eventually reaches a regime where significant 
``jumps'' appear in both $M_*$ and $M_{\rm BH}$. For most star-forming galaxies, 
this later phase is missing.

Much of the effort in this paper is to quantify the distinct phases and identify the 
underlying mechanisms driving the evolution during each phase and the transition to 
the next. In this paper, the main branches to be analyzed are selected according 
to the following criteria:
\begin{enumerate}[topsep=1pt,parsep=0pt,itemsep=1pt,leftmargin=12pt,label=(\roman*)]
    \item The root subhalo is a central subhalo at $z=0$.
    \item The root subhalo has $M_{\rm h} \geqslant 5\times10^{11} \Msun$, 
    $M_* \geqslant 10^{9} \Msun$, and has a SMBH in it. 
    \item The root galaxy is quenched.
    \item The branch can be traced up to snapshot 10 ($z\approx 7$). The histories of subhalos
    before this redshift are not used.
    \item The branch does not contain a ``backsplashed'' subhalo that
    temporarily falls into another halo, becomes a satellite, and 
    returns to being central after its halo mass exceeds one-third of
    the halo mass at $z=0$.
\end{enumerate}
The requirement for the subhalo to be a central (criterion i) and to have never 
been a satellite (criterion v) is to avoid the effect of the environment
which provides additional channels to quench the galaxy
\citep[see, e.g.][]{kauffmann2013ReexaminationGalacticConformity,
ayromlou2023PhysicalOriginGalactic,wang2023DissectTwohaloGalactic}.
The lower limit imposed on $M_*$ (criterion ii) 
ensures that the root subhalo has at least $\approx 1000$ stellar particles,
thus alleviating the artifacts due to numerical resolution.
The requirement of the complete history (criterion iv) 
and quenched state (criterion iii) ensures that none of the
four phases (see \S\ref{sec:def-phases}) is missed in the 
history, which simplifies the analysis and presentation of this paper.
The numbers of galaxies passing the above 5 criteria are 6291208, 8280, 1682, 
1594 and 1429, respectively. In addition, we have excluded {\revised two} 
galaxies that cannot be properly fitted by the function to be discussed 
in \S\ref{sec:def-phases}.
The conclusions of this paper thus apply only to quenched 
galaxies at $z=0$, for which we have checked that the above selection 
criteria do not lead to significant bias.
For star-forming galaxies, as shown in the top panel of Fig.~\ref{fig:sample-overall}, 
their histories appear similar to the quenched galaxies with some 
later phases truncated. Whether or not this similarity is quantitatively true,
how the evolution of star-forming galaxies in each phase and at each 
transition is different from that of quenched galaxies, 
and how SMBH feedback acts differently between star-forming and quenched galaxies
\citep[e.g.][]{fabianObservationalEvidenceActive2012,
duboisBlack2015,weinbergerSupermassiveBlackHoles2018}, remain to be 
open questions to be addressed further.

The lower panel of Fig.~\ref{fig:sample-overall} shows the main-branch 
histories in the $M_{\rm BH}$-$M_*$ plane for all the galaxies selected in 
our sample. The dashed red curve shows the median among all the histories.
We label the processes that drive the evolution of galaxies in different phases 
in this figure, and we will discuss them in detail in the following sections.

\section{Evolutionary Phases and Transitions} 
\label{sec:def-phases}

\begin{figure*}
\centering
\includegraphics[width=.9\textwidth]{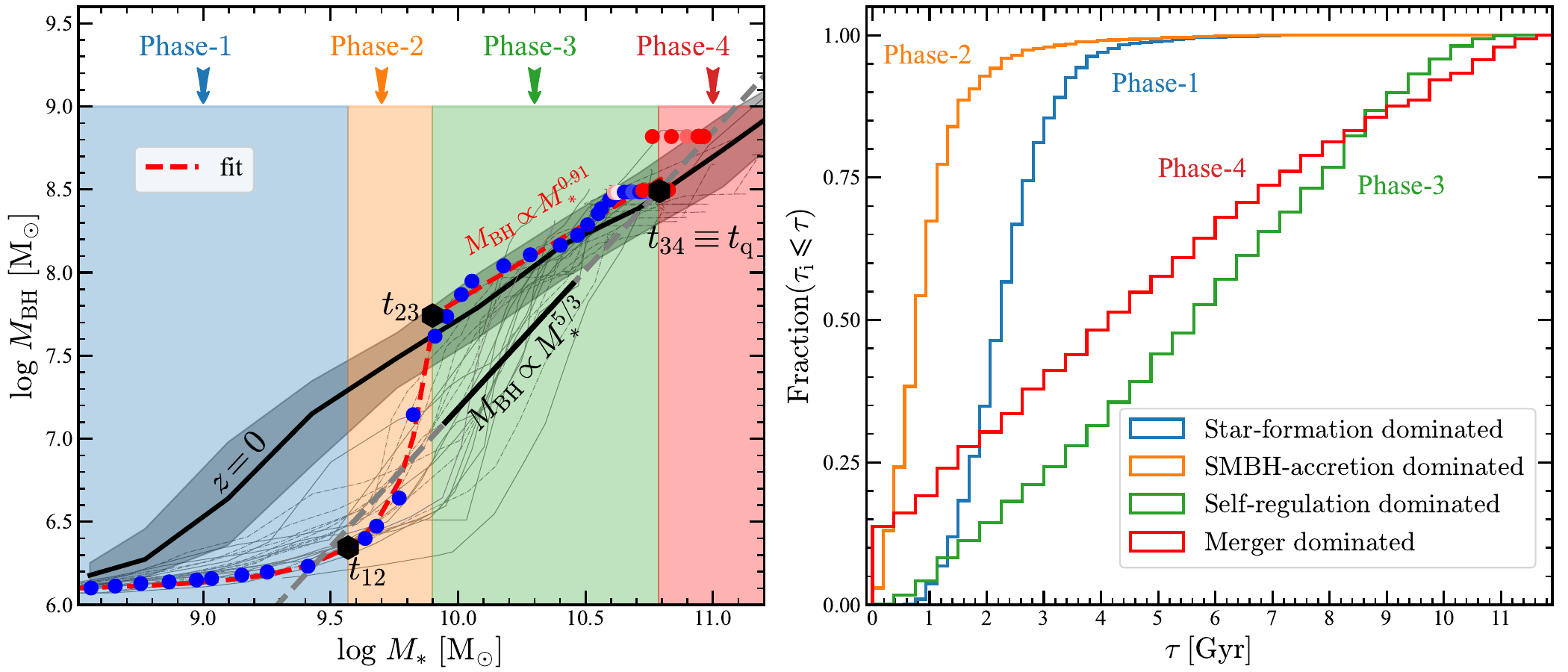}
\caption{
    The phases and transitions in the evolution histories of galaxies.
    \textbf{Left panel} shows the $M_{\rm BH}$-$M_{*}$ histories of 
    galaxies.
    \textbf{Grey curves} represent the simulated histories of 
    20 individual galaxies randomly selected from our sample 
    (see \S\ref{sec:data-and-sample}).
    Dots are the history of an example galaxy, colored blue (red) if 
    it is star-forming (quenched) at a snapshot.
    \textbf{Red dashed curve} shows the analytical fitting 
    of the history for this example.
    Four phases are defined according to the fitting function and are 
    labeled at the top of the panel.
    Three time points, $t_{\rm 12}$, $t_{\rm 23}$ and 
    $t_{\rm 34} \equiv t_{\rm q}$, are used to separate these phases
    (see \S\ref{sec:def-phases} for the fitting method and the definition
    of phases and transitions).
    \textbf{Black curve} shows the median $M_{\rm BH}$-$M_{*}$ relation 
    at $z=0$ computed using all galaxies in our sample, with
    the shaded area indicating the $16^{\rm th}$--$84^{\rm th}$ percentiles.
    \textbf{Short black line} indicates a power law, 
    $M_{\rm BH}\propto M_*^{5/3}$, to which the regulation of 
    thermal-mode AGN feedback in Phase-3 tends to drive the galaxy
    (see \S\ref{ssec:phase-3}). 
    \textbf{Right panel} shows the cumulative distribution of $\tau_i$, 
    the duration of Phase-$i$. 
}
\label{fig:fit-info} 
\end{figure*}

{\revised 
The similarity of the trajectories of the quiescent galaxies in 
Fig.~\ref{fig:sample-overall} indicates that a universal function can be used to fit
the evolution path of each quiescent galaxy in the $M_{\rm BH}$-$M_*$ plane.
To quantify the transitions and phases in the growth histories,
we fit all snapshots before the quenching time, $t_{\rm q}$, 
by the following piecewise continuous function:} 
\begin{equation} \label{eq:fitting-function}
y(x) = \begin{cases}
    \frac{(a+b)^3}{2} \left[ \frac{1}{(x-x_0-a)^2} -\frac{1}{a^2} \right] + y_0 \,, \
            & {\rm if}\ 0 \leqslant x < x_0  \,; \\ 
    k (x-x_0)+y_0  \,, 
            & {\rm if}\ x \geqslant x_0  \,,
    \end{cases}
\end{equation}
where $y = \log (M_{\rm BH}/\Msun)$, $x = \log (M_{\rm *}/\Msun)$, 
and $(x_0, y_0, k, a, b)$ are the five free parameters characterizing the evolutionary 
phases and transitions. The values of the free-parameters 
are obtained by minimizing the loss function,
$\chi^2=\sum_s(y_{{\rm sim},s}-y_{s})^2$, computed using 
the simulated value ($y_{\rm sim}$) and the value predicted by the 
fitting function ($y$) at every snapshot, $s$.
The optimization is performed using the Levenberg-Marquardt algorithm, which is 
designed for solving non-linear least squares problems. 
By design, the fitting function explicitly includes three transition points
at which the behavior of the galaxy in the $M_{\rm BH}$-$M_*$ plane
changes significantly. 
The first transition point is $x = x_0 - b$ at which ${\rm d}y/{\rm d}x = 1$. 
This marks the time when the growth of the SMBH begins to outpace the 
star formation. The second transition is at $x = x_0$, where the rapid 
growth in the SMBH mass ends. Given that TNG adopts a Bondi accretion 
upper-bound only by the Eddington limit, the growth of SMBH
during $x_0 - b < x < x_0$ can be very fast. 
This behavior can be captured by a value of $a$ close to zero.
For $x \geqslant x_0$, the function $y(x)$ is designed to be linear, describing a power-law 
relation, $M_{\rm BH} \propto M_*^k$.
The final transition is the ending point of the domain of the fitting function,
which, by design, is the quenching point of the galaxy.
The red dashed curve shows the best-fit function for an example galaxy 
in the left panel of Fig.~\ref{fig:fit-info}, 
with three black hexagons marking the three transition 
points (see below for the notations), respectively. 
The fitting captures well all the nonlinear features of the simulated points 
in this example. 
{\revised
Since mergers do not dominate the growth of $M_{\rm BH}$ and $M_*$ compared with 
the accretion of gas and in-situ star formation before quenching 
\citep[see e.g. figure 7 of][]{weinbergerSupermassiveBlackHoles2018}, 
the nuances introduced by mergers in the growth of galaxies in 
the $M_{\rm BH}$-$M_*$ plane do not significantly affect the fitting.
For reference, in Appendix~\ref{app:show_fit}, 
we quantify the goodness-of-fit for all galaxies in our sample. 
For most galaxies in our sample, the root-mean-square error is less than 
$0.1\dex$, and the maximum absolute error is less than $0.3\dex$, 
indicating that the fittings are reliable for 
individual galaxies. 
In addition, we have performed tests by tweaking the fitting function
and resampling the data points used in the fitting, 
and found that the identification of the transition points 
is robust against such changes.}

The three transition points break the evolution of a galaxy into four phases.
According to the physical processes to be introduced in \S\ref{sec:processes},
we name these phases as follows:
\begin{itemize}[topsep=4pt,parsep=4pt,itemsep=0pt,leftmargin=0pt]

\item {\em Phase-1: Star-formation dominated growth}.
This phase is characterized by a slow growth of the SMBH,  
with $M_{\rm BH}$ remaining close to the seeded mass, and a significant growth 
of $M_*$ due to active star formation. 

\item {\em Phase-2: SMBH-accretion dominated growth}.
This phase is featured by a rapid growth in $M_{\rm BH}$, by up to two orders 
of magnitude, and a modest growth in $M_*$, by a factor of about two to three.

\item {\em Phase-3: Self-regulation dominated growth}.
This is a period during which the masses of the SMBH and the stellar component 
grow at a comparable rate in logarithmic scales. 
This causes the galaxy to evolve towards a scaling relation of
$M_{\rm BH} \sim M_*^{5/3}$ (indicated by the black short line in the left panel of 
Fig.~\ref{fig:fit-info}) if it initially deviates from this relation.

\item {\em Phase-4: Merger dominated growth}.
This phase is characterized by the cease of in-situ star formation and 
SMBH growth. Consequently, the trajectory of a galaxy in the $M_{\rm BH}$-$M_*$ 
plane is stuck at around a point near the scaling relation at $z=0$, 
with some occasional ``jumps'' due to mergers.

\end{itemize}

For convenience, we denote the duration of Phase-$i$ by $\tau_i$, and 
the time point of the transition from Phase-$i$ to Phase-$j$ by $t_{ij}$. 
$\tau_1$ is defined to be the duration between $t_{\rm seed}$, the time when 
the SMBH is seeded, and $t_{12}$. $\tau_4$ is defined to be the duration between
$t_{\rm 34}$ and $z=0$.
For example, a galaxy stays in Phase-3 for a period of $\tau_3$,
transits at the time $t_{\rm 34}$ to Phase-4 and 
stays there for a period of $\tau_4$. The last transition point, $t_{34}$,
is defined to be the quenching time of the galaxy, and thus we use $t_{\rm q}$
for it interchangeably.

The right panel of Fig.~\ref{fig:fit-info} shows the cumulative distribution 
of the duration for each of the four phases. 
Phase-1 is short for most galaxies.
Phase-2 is even shorter due to the fast SMBH growth modeled by Bondi accretion, 
with $\approx 70\%$ galaxies staying there for $\lesssim 1 \Gyr$.
The last two phases have nearly uniform distributions of $\tau$ (seen 
from the nearly linear cumulative distributions), indicating that 
the quenching of a galaxy in TNG can happen at any time after the phase of 
rapid growth of SMBH.

\section{Processes Driving the Evolution}
\label{sec:processes}

In order to identify the underlying physical processes governing the distinct 
phases found phenomenologically by the curve fitting shown above, here we examine 
some physical properties in more detail, quantify their relations, demonstrate 
the conditions for their transitions, and discuss when and how they lead to the 
migration of a galaxy from one phase to the next. 

\begin{figure*}  
\centering
\includegraphics[width=.9\textwidth]{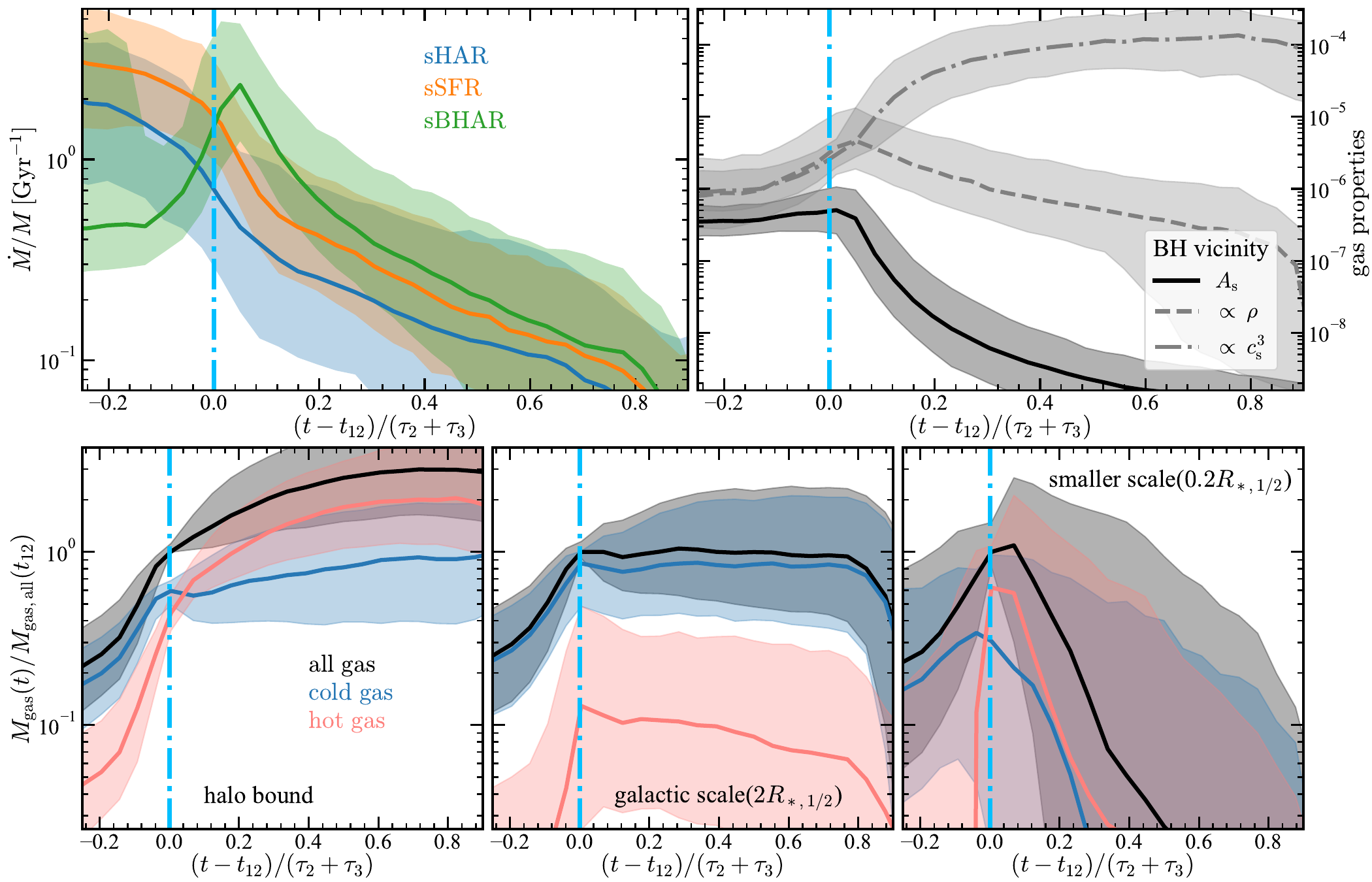}
\caption{ 
    Evolutionary history of various properties around the transition point
    $t_{12}$. Each property describing the history is averaged over 
    our sample (see \S\ref{sec:data-and-sample}) and shown as a function
    of time, measured with respect to $t_{12}$ and scaled by the duration
    $\tau_2 + \tau_3$. The shaded area around each curve shows the 
    $16^{\rm th}$--$84^{\rm th}$ percentiles.
    Each panel shows a set of related properties.
    {\bf Top left} panel shows the specific growth rates of 
    of $M_{\rm h}$ (blue), $M_*$ (orange) and 
    $M_{\rm BH}$ (green).
    {\bf Top right} panel shows the gas properties involved in the definition
    of Bondi accretion rate (see Eq.~\ref{eq:mbh-dot-phase-1} and 
    \ref{eq:mbh-dot-numeric}) in the vicinity of the central SMBH.
    Three {\bf bottom panels} show the gas masses defined at different scales:
    bound to the subhalo ({\bf left}), within $2 R_{\rm *,1/2}$ 
    ({\bf middle}), and within $0.2 R_{\rm *,1/2}$ ({\bf right}). Total,
    cold ($ \leqslant 10^5 \Kelvin$), and hot ($ > 10^5 \Kelvin$) gas masses 
    are shown by black, blue and red curves, respectively, all scaled by 
    total gas mass at $t_{12}$. 
}
\label{fig:phase-1-mass-history}
\end{figure*}

\begin{figure}
    \centering
    \includegraphics[width=.45\textwidth]{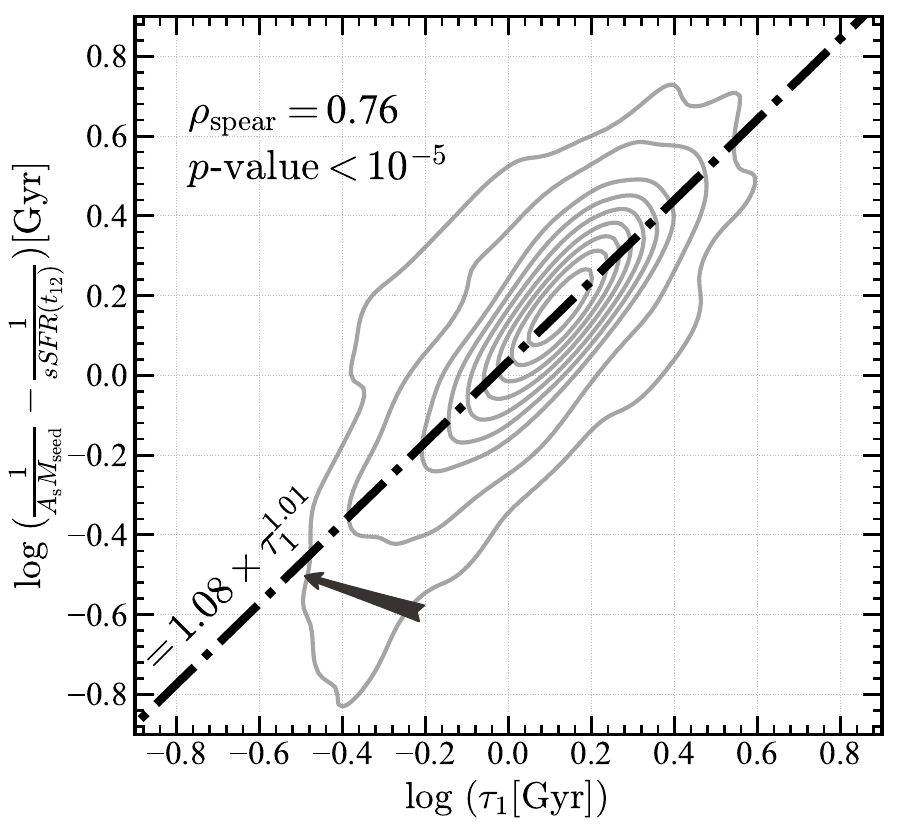}
    \caption{
        Relation between the estimated value and the simulated value of 
        the Phase-1 duration ($\tau_{1}$). 
        {\bf Grey curves}, from inner to outer, are contours enclosing 
        10\% -- 95\% of the sample. 
        {\bf Dashed line} is a power-law fitting (in linear scale) of the 
        relation. The normalization and power index are both close to 1, 
        indicating that the estimated value well approximates the simulated value.
        See \S\ref{ssec:phase-1} for the details.
    }
    \label{fig:phase-1-time-estimate} 
\end{figure}

\subsection{Phase-1: Star-formation dominated growth}
\label{ssec:phase-1}

The growth of a galaxy in the first phase is featured by a small mass of the 
SMBH. In TNG, a seed of SMBH with a mass of $M_{\rm BH} = 1.18 \times 10^6 \Msun$ 
is placed into a galaxy when the mass of its host halo exceeds 
$7.38 \times 10^{10} \Msun$ \citep{weinbergerSupermassiveBlackHoles2018}.
The subsequent growth of the SMBH is modeled by the Bondi-Hoyle accretion
\citep[][thereafter Bondi accretion]{weinbergerSimulatingGalaxyFormation2017}.
As the accretion rate of the SMBH modeled in this way scales as $M_{\rm BH}^2$, 
the seed thus determines the initial sBHAR.
In TNG, the seed mass is small enough so that the sBHAR is initially negligible,
which, together with a high sSFR at high redshift \citep[e.g.][see their figure 4]{habouzitSupermassiveBlackHoles2021},
drives a horizontal trajectory in the $M_{\rm BH}$-$M_*$ plane for the galaxy
(see the left panel of Fig.~\ref{fig:fit-info}). 
The low accretion rate of SMBH also implies that its feedback effect is small,
and thus the star formation in the galaxy is regulated only by the 
stellar feedback. A comparison of the energy release rates via different feedback channels
by \citet[see their figure 1]{weinbergerSupermassiveBlackHoles2018} indeed shows
the domination of stellar feedback in the early time.
The first phase is thus referred to as the star-formation dominated phase
in this paper. 

The top left panel of Fig.~\ref{fig:phase-1-mass-history} shows the specific
growth rate of different masses ($M_{\rm h}$, $M_*$ and $M_{\rm BH}$) as a 
function of time. Each rate is averaged over the sample defined in \S\ref{sec:data-and-sample}.
To see the growth of the whole sample during the early phases, the time is 
measured with respect to $t_{\rm 12}$, the ending point of 
Phase-1, and scaled by $\tau_2 + \tau_3$, the total time of the following 
two phases. The exclusion of $\tau_1$ from the scaling factor is to avoid the ambiguity 
caused by the seeding time, and the inclusion of $\tau_3$ accounts for the fact
that $\tau_2$ is too short. During Phase-1, the sSFR follows the sHAR tightly, indicating 
that the formation of the stellar contents is driven by the gas accretion along 
with the halo growth and that the $M_*$-$M_{\rm h}$ relation has been well established 
before the seeding of SMBH. The sBHAR is significantly lower compared to the rates 
of the halo and galaxy, indicating that the early SMBH growth can deviate from
the $M_{\rm BH}$-$M_*$ relation at $z \approx 0$ in both TNG and observations
\citep[e.g.][]{Kormendy2013,greeneIntermediateMass2020,grahamAppreciatingMergersUnderstanding2023,
grahamReading2023,habouzitSupermassiveBlackHoles2021,zhuangEvolutionaryPathsActive2023}.
Thus, the growth of SMBH in Phase-1 has to be understood by combining
the established environment of the galaxy and the halo, and the specific 
strategy in the seeding of SMBH. 

To quantify the growth of SMBH in Phase-1, we formulate the Bondi accretion rate as
\begin{equation} \label{eq:mbh-dot-phase-1}
    \dot{M}_{\rm BH} = A_{\rm s} M_{\rm BH}^2 \,,
\end{equation}
where the normalization factor 
\begin{align} \label{eq:mbh-dot-numeric}
    A_{\rm s} \equiv &\ \frac{4\pi G^2 \rho_{\rm gas}}{c_{\rm s}^3} =  
    \left(\frac{1}{1.4\Gyr}\right) \times \nonumber\\ 
    & \left(\frac{c_{\rm s}}{10 \kms}\right)^{-3}  
    \left(\frac{n_{\rm gas}}{0.1 \perccm}\right)
    \left(\frac{1}{10^6 \Msun}\right)
    \,,
\end{align}
$\rho_{\rm gas}$ and $c_{\rm s}$ are the gas density and effective sound speed, 
respectively, within a smoothing kernel around the SMBH,
$n_{\rm gas}$ is the number density of gas molecules, assuming a mean molecular
weight of $\mu = 1.2$.
Thus, $A_{\rm s}$ depends only on the gas environment around the SMBH and,
by construction, is a function of the specific entropy, which is shaped by 
the dissipation and heating processes.
The bottom three panels of Fig.~\ref{fig:phase-1-mass-history} show the evolution 
of gas mass within different spatial extents.
The total mass, and those in hot and cold phases, are shown by curves 
with different colors. Within $2R_{\rm *,1/2}$, cold gas dominates the total gas 
mass during Phase-1, suggesting that supernova feedback alone cannot quench the 
star formation.
This is consistent with the cooling rate of non-star-forming cells computed by 
\citet[see their figure 1]{weinbergerSupermassiveBlackHoles2018}, which always balances
the energy output from stellar feedback in the early time.
In the innermost region ($r \lesssim 0.2 R_{\rm *,1/2}$) during Phase-1, 
the cold component also dominates the gas mass, which, together with the high 
gas density in this region \citep[e.g.][]{bowerDark2017b}, implies that 
the early growth of SMBH is fed mainly by star-forming gas.
The quantitative state of star-forming gas cells, however, depends on the 
subgrid recipes assumed by TNG, which cannot be estimated analytically. 
We therefore directly compute the state variables from the simulation data.
The gas density, $\rho_{\rm gas}$, is averaged by an SPH kernel 
\citep[e.g.][]{weinbergerSimulatingGalaxyFormation2017} over a
sphere centered on the SMBH and enclosing approximately $256$ cells 
(defined by the radius {\tt BH\_Hsml} in the particle data of TNG).
The effective sound speed, $c_{\rm s}$,
which includes the contribution from the Alfv\'en speed of the magnetic field,
is obtained by inserting Eq.~\eqref{eq:mbh-dot-phase-1} using the values 
of $\rho_{\rm gas}$ estimated as above, and $M_{\rm BH}$ and $\dot{M}_{\rm BH}$ 
read from the simulation data. 
The evolutions of $A_{\rm s}$, $\rho_{\rm gas}$ and $c_{\rm s}$ are 
shown in the top right panel of Fig.~\ref{fig:phase-1-mass-history}.
An interesting fact is the constant value of $A_{\rm s}$ during Phase-1,
which implies that the gas processes around the SMBH are effectively 
adiabatic.
We note that this is an outcome of the specific choice of TNG on the effective 
equation of state (eEOS) for star-forming cells and the subgrid model 
for the magnetic field \citep[e.g.][]{springel2003CosmologicalSmoothedParticle,
vogelsbergerModel2013,pillepichSimulatingGalaxyFormation2018a}.

Using the fact that $A_{\rm s}$ is roughly constant in Phase-1, we integrate 
Eq.~\eqref{eq:mbh-dot-phase-1} and obtain
\begin{equation} \label{eq:mbh-t-phase-1}
    \frac{M_{\rm BH}(t)}{M_{\rm seed}} = 
    \frac{1}{1 - A_{\rm s} M_{\rm seed} \cdot (t - t_{\rm seed}) } \,,
\end{equation}
where $M_{\rm seed} \equiv M_{\rm BH}(t_{\rm seed})$. This solution becomes 
divergent at $t - t_{\rm seed} = (A_{\rm s}M_{\rm seed})^{-1}$, which 
sets the upper limit of $\tau_1$.
By our definition, the ending point of Phase-1 is when sBHAR equals sSFR. 
Substituting this definition into 
Eqs.~\eqref{eq:mbh-dot-phase-1} and \eqref{eq:mbh-t-phase-1}, we obtain the duration of 
Phase-1 as 
\begin{equation} \label{eq:tau-1-estimate}
    \tau_{\rm 1} = \frac{1}{A_{\rm s} M_{\rm seed}} - \frac{1}{\rm sSFR} 
    \leqslant \frac{1}{A_{\rm s} M_{\rm seed}} \,.
\end{equation}
This result implies that, the duration of the phase when star formation 
outpaces SMBH growth is determined by 
(i) the seeding strategy of SMBH in the simulation, or the formation of SMBH seed in the real Universe;
(ii) the specific entropy of gas surrounding the SMBH seed, 
namely, the boundary conditions provided by halo accretion, star formation and 
stellar feedback; (iii) the definition of the end of the phase.
The upper limit of $\tau_1$, $(A_{\rm s} M_{\rm seed})^{-1}$, 
is determined by only (i) and (ii).

Fig.~\ref{fig:phase-1-time-estimate} shows the distribution of $\tau_{1}$  
directly obtained from the fitting function described in \S\ref{sec:def-phases}, 
in comparison with the value estimated by Eq.~\eqref{eq:tau-1-estimate}.
To avoid large fluctuations at a specific snapshot, we use the average 
$A_{\rm s}$ during Phase-1. 
The contour in the figure shows the distribution of SMBHs and the dot-dashed line 
shows a power-law fitting. 
The normalization and power index of the fitting are both close to 1, 
indicating that the estimation of $\tau_{1}$ by Eq.~\eqref{eq:tau-1-estimate} 
is accurate. 
Thus, {\revised from Eq.~\eqref{eq:tau-1-estimate}}, the transition from star-formation 
dominated growth to SMBH-accretion
dominated growth, {\revised with a given definition of the transition}, 
is a combined consequence of the subgrid model adopted 
by TNG: the low-mass seeds, the Bondi accretion rate, and the eEOS of 
star-forming gas.

A critical task in hydrodynamical simulations is to implement different
recipes of SMBH seeding and growth, to embed SMBHs in different
environments, and to understand their observational consequences.
For example, \citet{bowerDark2017b} analytically estimated the condition 
for the triggering of rapid SMBH growth 
(the transition at $t_{\rm 12}$ in our terminology), 
and tested it in the EAGLE simulation with a list of model variants. 
As EAGLE is different from TNG in its hydrodynamic solver and the 
thermal-mode stellar feedback, the results and the interpretations can 
be different.
Indeed, their results suggested that the transition at $t_{12}$ is caused 
by the transition of the effect of stellar feedback, which is initially 
effective in expelling gas from the galaxy by the buoyant outflow, but then 
becomes ineffective when the halo grows massive and the gas environment
of the galaxy becomes hot. The transition is thus controlled by a 
critical halo mass, which is suggested to be $\sim 10^{12}\Msun$ and
depend moderately on redshift. A follow-up study by 
\citet{mcalpineRapidGrowthPhase2018} based on EAGLE instead suggested
that the transition occurs at a nearly constant virial temperature of
$\approx 10^{5.6}\Kelvin$.
Other works also suggested that active star formation can expel gas from the 
vicinity of the central SMBH, reduce the gas density and suppress the accretion 
of the SMBH \citep[e.g.][]{habouzitBlossomsBlackHole2017,
angles-alcazarBlackHolesFIRE2017,
hopkinsWhyBlackHoles2021a,mcalpineRapidGrowthPhase2018,
byrneStellarFeedbackregulatedBlack2023}.
These models thus create environments for the growth of SMBH 
in the star-formation dominated phase that are different from that in TNG,
and are expected to produce $\rho_{\rm gas}(t)$ and $c_{\rm s}(t)$ curves that are different
from the smooth, slowly increasing curves shown in the top right panel of 
Fig.~\ref{fig:phase-1-mass-history}.
The evolution of $A_{\rm s}$ may also be changed, which can lead to 
an evolution equation of $M_{\rm BH}$ different from our 
Eq.~\eqref{eq:mbh-t-phase-1}.

After $t_{\rm 12}$, the rapid growth of the SMBH starts together with 
strong thermal-mode feedback.
Using the estimation given by \citet[see their \S2.4]{weinbergerSupermassiveBlackHoles2018}, 
the ratio of the energy output rate from thermal-mode AGN and star formation 
for the metal-enriched ($Z \gg 0.002$) gas is 
\begin{equation} \label{eq:energy-ratio}
    \frac{\dot{E}_{\rm AGN, thermal}}{\dot{E}_{\rm SF}} 
    \approx 3 \times 10^3 \frac{\dot{M}_{\rm BH}}{\dot{M}_*}\,.
\end{equation}
Therefore, once $M_{\rm BH}/M_* \gtrsim 10^{-3}$, the cumulative energy output 
from AGN dominates over that from star formation, and the galaxy starts to 
migrate from the supernova-regulated phase to the AGN-regulated phase.
The example galaxy in Fig.~\ref{fig:fit-info} shows that 
$M_{\rm BH}/M_*$ indeed reaches this critical value at around $t_{\rm 12}$,
thus supporting our definition of this transition point and 
the interpretation of it as a transition to the SMBH-accretion 
dominated phase.

\subsection{Phase-2: SMBH-accretion dominated growth}
\label{ssec:phase-2}

\begin{figure} 
    \centering
    \includegraphics[width=.45\textwidth]{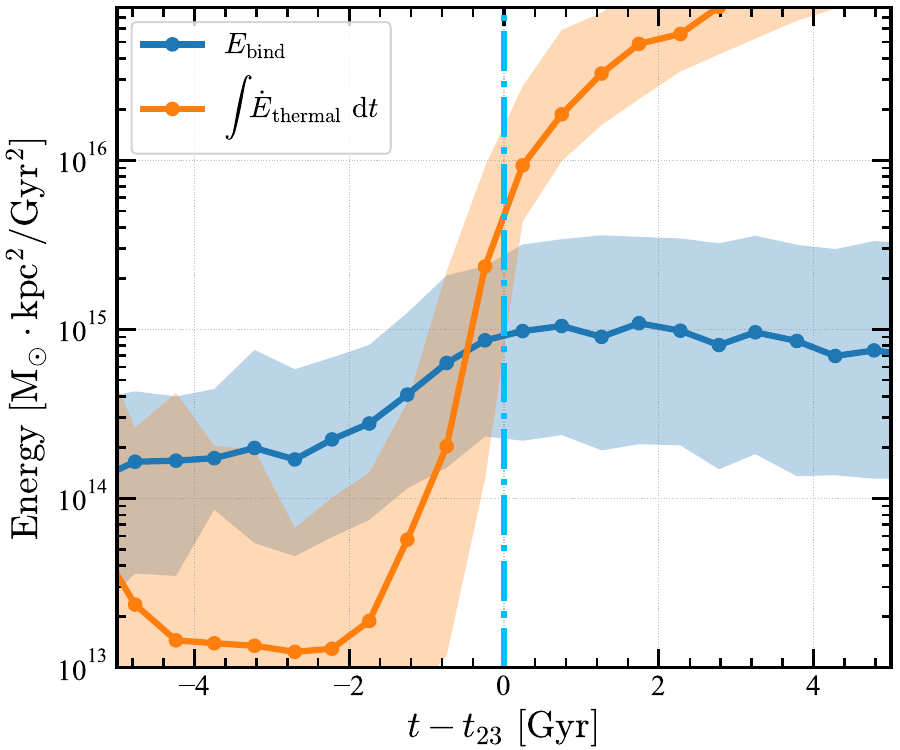}
    \caption{
        Evolution of energies at around the transition point $t_{23}$.
        {\bf Blue curve} shows $E_{\rm bind}$, the binding energy of the gas
        within the galaxy. {\bf Orange curve} shows $E_{\rm agn}$, 
        the cumulative feedback energy from AGN.
        Both are shown as a function of time, measured with respect to $t_{23}$.
        See \S\ref{ssec:phase-2} for the details.
    }
    \label{fig:gas-unbinding} 
\end{figure}

\begin{figure*} 
\centering
\includegraphics[width=.9\textwidth]{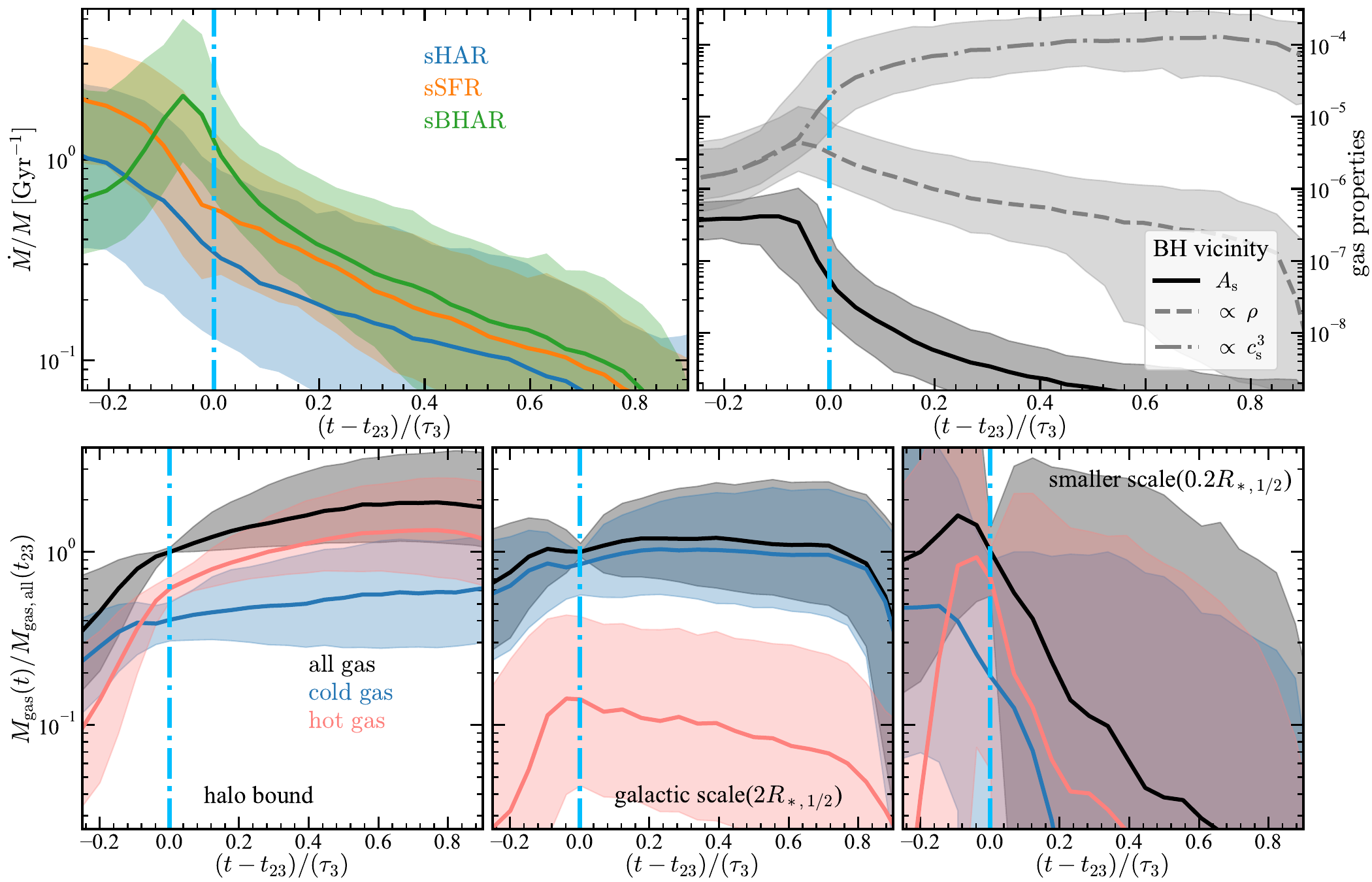}
\caption{
    Similar to Fig.~\ref{fig:phase-1-mass-history}, but here the 
    time is measured with respect to $t_{23}$ and scaled by $\tau_3$, 
    so that the evolution of galaxies around the transition from 
    SMBH-accretion dominated growth to self-regulation dominated 
    growth is clear.
} 
\label{fig:phase-2-mass-history}
\end{figure*} 

The transition point, $t_{\rm 12}$, by definition, is the time when the 
sBHAR becomes larger than the sSFR.
As shown in the top left panel of Fig.~\ref{fig:phase-1-mass-history}, 
the sBHAR and sSFR cross each other at this point. 
In the $M_{\rm BH}$-$M_*$ plane, the trajectory makes a transition
at $t_{12}$ from nearly horizontal to nearly vertical, as seen 
from the example shown in Fig.~\ref{fig:fit-info}.
After this point, the sBHAR continues to increase, reaches a peak, and then 
declines to a rate comparable to the sSFR. In contrast, the sSFR always declines 
with time and remains below the sBHAR during Phase-2.
The obvious reason for the fast growth of the SMBH in Phase-2 is the Bondi accretion 
implemented by TNG. For a given gas environment, the BHAR is $\propto M_{\rm BH}^2$,
so that
\begin{equation} \label{eq:bondi-growth}
    \dv{\ln M_{\rm BH}}{t} \propto \exp(\ln M_{\rm BH})\,.
\end{equation} 
Thus, the growth of SMBH in the logarithmic scale is exponential.
This growth, however, can only last for a short period before 
the properties of the surrounding gas are changed significantly by the AGN feedback.
The cumulative distribution in the right panel of Fig.~\ref{fig:fit-info} 
shows that the duration of Phase-2 is the shortest among the four phases,
with a median of $\approx 1\Gyr$.

To derive a general condition for the ending of Phase-2, 
we start with the equation of energy conservation for the gas within the 
galaxy:
\begin{equation} \label{eq:energy-conservation}
    \dd{E_{\rm gas}} = -\dd{E_{\rm cool}} + \dd{E_{\rm fdbk}}\,,
\end{equation}
where $E_{\rm gas} \equiv E_{\rm in} + E_{\rm k} + E_{\rm v}$ is {\revised the total 
energy of the gas}, defined as the sum of the internal, kinetic, and
gravitational potential energies, $E_{\rm cool}$ is the energy loss due to 
radiative cooling, and $E_{\rm fdbk}$ is the energy input from feedback.
The ending of fast SMBH accretion requires a significant reduction in the 
amount of gas that can fuel the SMBH. To be achieved, the gas of the entire galaxy, in 
particular that in the inner region, has to be significantly affected by the feedback, 
because otherwise cold gas clumps that can form in the galaxy will sink into 
the center and fuel the SMBH.
Motivated by the flat curve of the total gas mass 
shown in the bottom center panel of Fig.~\ref{fig:phase-1-mass-history}
after $t_{12}$, we express the gas mass at time $t$ around $t_{\rm 23}$ as 
\begin{equation}
    M_{\rm gas}(t) - M_{\rm gas}(t_{\rm 23}) 
    \approx \dot{M_{\rm}} (t_{\rm 23}) (t - t_{\rm 23}) = 0\,,
\end{equation}
With this, the conservation equation \eqref{eq:energy-conservation} can be 
rewritten as
\begin{align} \label{eq:energy-conservation-eff}
    M_{\rm gas} \dd{\epsilon_{\rm gas}} 
    &= -\dd{E_{\rm cool}} + \dd{E_{\rm fdbk}}  \nonumber\\
    &= \eta_{\rm eff} \dd{E_{\rm fdbk}} \,,
\end{align}
where $\epsilon_{\rm gas} \equiv E_{\rm gas}/M_{\rm gas}$ is the specific 
energy of gas, and $\eta_{\rm eff}$ is the effective fraction of the 
feedback energy that is coupled to the gas after subtracting radiative cooling.
The time integration of the left-hand side of 
Eq.~\eqref{eq:energy-conservation-eff} until the gas becomes unbound
is exactly the binding energy, and we use the virial theorem
\citep[e.g.][]{terrazasRelationship2020} to estimate it:
\begin{equation} \label{eq:e-bind}
    M_{\rm gas} \int \dd{\epsilon_{\rm gas}} = E_{\rm bind}
    \approx
    \frac{1}{2}  \sum_{\rm gas} m_{\rm gas}\phi_{\rm gas} \,.
\end{equation}
Here, the summation is over gas cells within the radius of the galaxy, defined 
as $2 R_{\rm *,1/2}$ (see \S\ref{sec:data-and-sample}), 
and $\phi_{\rm gas}$ is the gravitational potential at the location 
of a gas cell.
The integration of the right-hand side of Eq.~\eqref{eq:energy-conservation-eff} 
is the cumulative value accounting for both feedback and cooling. 
Because AGN feedback dominates the feedback energy after $t_{12}$ 
(see Eq.~\ref{eq:energy-ratio} and the texts below it),
we can obtain $E_{\rm fdbk}$ as
\begin{align} \label{eq:e-fdbk}
    E_{\rm fdbk} & \approx E_{\rm agn} =  \int \dot{E}_{\rm thermal}\dd{t} \nonumber\\
    & = \int \eta_{\rm thermal} \dot{M}_{\rm BH} c^{2} \dd{t}
    \approx \eta_{\rm thermal} M_{\rm BH} c^{2}\,,
\end{align}
where $\eta_{\rm thermal} = 0.02$ is the efficiency of the thermal-mode feedback 
\citep{weinbergerSupermassiveBlackHoles2018}. {\revised
To exclude the contribution from mergers to $\dot{M}_{\rm BH}$ in the estimation
of $E_{\rm agn}$, we take the field {\tt BH\_CumEgyInjection\_QM} 
from the catalog of TNG to obtain the feedback energy injected by the BH 
in the thermal mode.} 
The radiative cooling is dominated by star-forming gas cells because of their 
high density and the sensitive dependence of the cooling rate on gas density.
Unfortunately, star-forming gas cells are not resolved by the simulation,
which prevents us from computing the cooling rate directly. 
\citet{weinbergerSupermassiveBlackHoles2018} suggested an alternative way to estimate
the cooling loss by subtracting the feedback energy deposited to the star-forming
gas cells. This gives a value of $\eta_{\rm eff} \approx 10\%$--$20\%$ 
for TNG, as shown in their figure 1.
Substituting Eqs.~\eqref{eq:e-bind} and \eqref{eq:e-fdbk} into 
Eq.~\eqref{eq:energy-conservation-eff}, we can 
obtain the condition for the ending of the fast SMBH accretion as 
\begin{equation} \label{eq:condition-t-23}
    E_{\rm agn} \gtrsim \eta_{\rm eff}^{-1} E_{\rm bind}\,,
\end{equation}
where $\eta_{\rm eff}^{-1} \approx 5$--$10$ for TNG.

Fig.~\ref{fig:gas-unbinding} shows the change of $E_{\rm bind}$ and $E_{\rm agn}$ as 
a function of time around the transition point $t_{\rm 23}$. 
As expected, the total binding energy $E_{\rm bind}$ stops increasing
at $t_{\rm 23}$, indicating that part of the gas that flows from the halo to the galaxy
is ejected by feedback. The feedback energy $E_{\rm agn}$ from AGN starts to rise 
rapidly at $\approx 2\Gyr$ prior to $t_{\rm 23}$, consistent with the duration 
$\tau_{2}$ of Phase-2 shown in Fig.~\ref{fig:fit-info}. 
$E_{\rm agn}$ surpasses $E_{\rm bind}$ at $\approx 0.2 \Gyr$ prior to $t_{\rm 23}$, 
and reaches $\approx 7\times E_{\rm bind}$ at $t_{\rm 23}$. Thus, the condition given 
by equation \eqref{eq:condition-t-23} is satisfied, and the fast SMBH accretion is stopped.

The details of the gas ejection by AGN feedback depend on the structure
and thermal state of the gas, the coupling of the feedback energy to the gas, 
and the transfer of energy among gas cells.
In TNG, the energy of the thermal-mode feedback is directly deposited to the gas cells 
surrounding the SMBH and transferred to the outer region via hydrodynamical 
processes \citep{weinbergerSimulatingGalaxyFormation2017}.
Thus, the impact of the thermal-mode feedback is inside-out: it significantly reduces 
the amount of gas surrounding the SMBH, but has smaller effects on outer regions.
The three bottom panels of Fig.~\ref{fig:phase-2-mass-history} show the change of 
the gas mass within different scales.
As expected, the gas mass on halo scales, for both the cold and hot components, 
continues to increase. The mass of the galactic-scale gas stops growing at around 
$t_{\rm 23}$. Within $0.2 R_{\rm *,1/2}$, the cold gas is heated and ejected 
by the AGN feedback, leading to a reduction of the total gas mass 
after $t_{\rm 23}$. The temperature of the gas within the galaxy is 
raised, but the density drops, as shown in the top right panel of
Fig.~\ref{fig:phase-2-mass-history}.
The significant inside-out change of the gas profile during Phase-2 leads to 
the cease of SMBH accretion, but only a moderate reduction in
star formation, as seen from the top left panel of 
Fig.~\ref{fig:phase-2-mass-history}. 
The evolutionary path of the galaxy in the 
$M_{\rm BH}$-$M_*$ plane, as shown by the example in Fig.~\ref{fig:fit-info}, 
is then pulled back from nearly vertical to a shallower slope. Note that the 
results obtained here are consistent with the conclusion of \citet{terrazasRelationship2020} 
that the thermal-mode AGN feedback primarily regulates the growth of the SMBH, 
but cannot quench the star formation. Instead, the growth of the galaxy 
enters into a self-regulated phase, as discussed below.

\subsection{Phase-3: Self-regulation dominated growth}
\label{ssec:phase-3}  

As the impact volume of AGN feedback expands to the entire galaxy, the gas 
in the galaxy enters into a regime that is dominated by the AGN feedback. 
As Bondi accretion depends on the gas properties around the SMBH, the growth 
of SMBH is self-regulated by its own feedback. The situation is similar to 
that reported by \citet[][see their figure 4]{zhuangEvolutionaryPathsActive2023} 
based on the reconstruction of the star-formation and SMBH-accretion histories
for a sample of observed broad-line AGNs. By tracing the evolution of 
individual systems, we are now able to derive the scaling relations between 
masses and the conditions to establish them.

\subsubsection[]{The self-regulated growth}
\label{ssec:smbh-self-regularization}

\begin{figure}[htb]
    \centering
    \includegraphics[width=.45\textwidth]{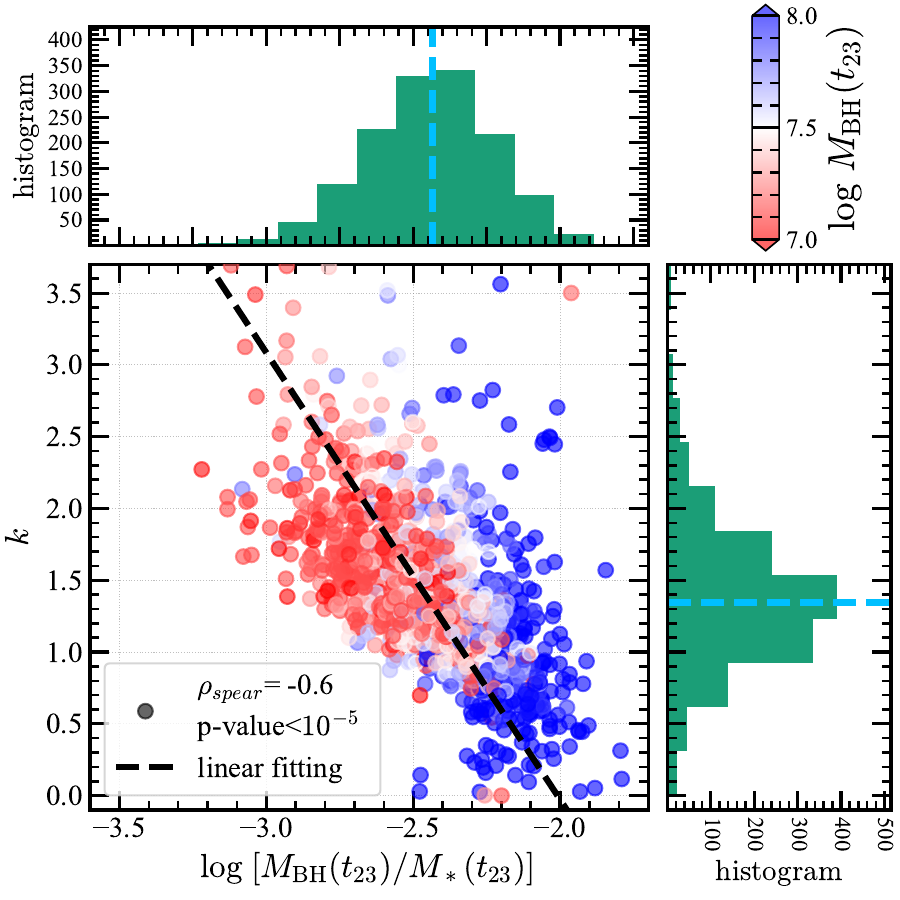}
    \caption{
        The relation between the power-law index, $k$, of the $M_{\rm BH}$-$M_*$ 
        history during Phase-3 and the ratio, $M_{\rm BH}/M_*$, at $t_{23}$, 
        the beginning of Phase-3.
        Dots in the {\bf bottom left} panel represent individual galaxies,
        color-coded by the value of $M_{\rm BH}$ at $t_{23}$.
        The Spearman correlation coefficient and the $p$-value are indicated
        in the label.
        {\bf Black dashed line} shows a linear fit, which gives a slope of about
        $-3.0$. 
        {\bf Top left} and {\bf bottom right} panels show the two marginal
        distributions, respectively. See \S\ref{ssec:smbh-self-regularization}
        for the details of how this figure implies a self-regulated growth.
    }
    \label{fig:phase-3-self-regularization} 
\end{figure}

The self-regularization nature can be seen from Eq.~\eqref{eq:condition-t-23}
by considering a perturbation on the gas properties and the consequence of 
such perturbation on the feedback energy ($E_{\rm agn}$) and the 
binding energy ($E_{\rm bind}$). If gas inflow is temporarily enhanced by, e.g.
the fluctuations of halo assembly or the disk instability, Bondi accretion 
will be accelerated, leading to a trajectory in the $M_{\rm BH}$-$M_*$ 
plane that resembles the one in Phase-2 and producing an over-massive SMBH.
The feedback output then rises, leading to an inequality,
$E_{\rm agn} > \eta_{\rm eff}^{-1} E_{\rm bind}$. The gas surrounding the SMBH 
will then be evacuated and the growth of SMBH will be stalled.
Inversely, if the gas inflow is suppressed by, e.g. the temporary reduction of 
halo accretion or the consumption of gas by star formation, Bondi accretion will be 
stopped, leading to a shallower trajectory in the $M_{\rm BH}$-$M_*$
plane. The inequality will be inverted as 
$E_{\rm agn} < \eta_{\rm eff}^{-1} E_{\rm bind}$,
and so the gas will be accumulated again, and the growth of the SMBH will be
accelerated. The discussion here implies that the evolution
of the galaxy and its SMBH closely follows a trajectory defined by the 
equilibrium equation $E_{\rm agn} = \eta_{\rm eff}^{-1} E_{\rm bind}$.

To convert this prediction into an explicit relation, we use the virial 
theorem, and approximate the binding energy as 
$E_{\rm bind} \sim M_{\rm g} V_{\rm g}^2$ {\revised at the galaxy scale}. 
Combined with the feedback energy
given by Eq.~\eqref{eq:e-fdbk}, we obtain 
\begin{align} \label{eq:phase-3-scaling}
    \eta_{\rm eff} \eta_{\rm thermal} M_{\rm BH}
    & \propto M_{\rm g} V_{\rm g}^2  
    \propto f_{\rm gas} M_{\rm h}^{5/3}       \nonumber \\
    & \propto f_{\rm gas} f_{\rm *}^{-5/3} M_*^{5/3}\,,
\end{align}
where we have defined $f_{\rm gas} = M_{\rm g}/M_{\rm h}$ 
and $f_* = M_*/M_{\rm h}$, and used the self-gravitation condition 
$V_{\rm g} \approx V_{\rm h} \sim M_{\rm h}^{1/3}$.
Thus, the equilibrium equation can be written as
\begin{equation}
    M_{\rm BH} \sim f M_*^{5/3}    \,,
\end{equation}
where the scaling factor 
$f \equiv \eta_{\rm eff}^{-1} \eta_{\rm thermal}^{-1} f_{\rm gas} f_*^{-5/3}$.
The values of $\eta_{\rm eff}$ and $\eta_{\rm thermal}$ depend on the 
implementation of the feedback model in the simulation, as described in
\S\ref{ssec:phase-2}. One interesting feature of $f_*$, 
as suggested by observations \citep[e.g.][]{Yang2007,yang2012EVOLUTIONGALAXYDARK,behrooziUniverseMachineCorrelationGalaxy2019}, 
is the flattening at the critical mass, $M_{\rm h,c} \approx 10^{12} \Msun$ 
(or $M_{\rm *,c} \approx 10^{10.5} \Msun$), 
where $f_*$ appears to be independent of halo mass to the first order. 
The gas fraction, $f_{\rm gas}$, is also expected to peak at this 
critical mass, as suggested by \citep{terrazasRelationship2020,voitBlackHoleGrowth2024}.
Thus, at around $M_{\rm *,c}$, all these coefficients are roughly constant, 
with only moderate redshift dependence
introduced by the relation between $M_{\rm h}$ and $V_{\rm g}$.
Consequently, the equilibrium equation predicts a power-law relation, 
$M_{\rm BH} \propto M_*^{5/3}$. At lower or higher mass scales, $f_{*}$ and $f_{\rm gas}$ 
cannot be viewed as a constant, and the power-law index may deviate from $5/3$.
Note that the prediction of the scaling relation here is consistent with the 
semi-analytical model of \citet{moTwophaseModelGalaxy2024} and the 
observationally calibrated toy model of \citet{Hong2023}, although they focus on the formation of 
bulge component and quenched galaxies, respectively.

The example shown by the red dashed curve in Fig.~\ref{fig:fit-info} demonstrates the regulation 
over the equilibrium scaling relation. At $t_{\rm 23}$, the SMBH mass
is slightly overshot relative to the power-law scaling with index $\gamma=5/3$. 
During Phase-3, the galaxy is pulled back towards the scaling, and 
finally comes to a region in the $M_{\rm BH}$-$M_*$ plane where other
galaxies (shown by grey curves) converge. To statistically quantify the self-regulation nature, 
in Fig.~\ref{fig:phase-3-self-regularization}, we show $k$, 
the power-law index of the $M_{\rm BH}$-$M_*$ relation during 
Phase-3 obtained from the best-fit function (Eq.~\ref{eq:fitting-function}), 
and $M_{\rm BH}(t_{23})/M_*(t_{23})$, the ratio of SMBH mass and stellar mass 
at the beginning of Phase-3, for individual galaxies, as well as the
marginal distributions.
The distribution of $M_{\rm BH}(t_{23})/M_*(t_{23})$ has a median of 
$\approx 10^{-2.4}$, approaching the typical value of galaxies at $z=0$.
The distribution of $k$ has a median of $\sim 1.35$, 
smaller than the value of $5/3$ that Eq.~\eqref{eq:phase-3-scaling}
would imply if both $f_{\rm gas}$ and $f_*$ are constant.
A strong anti-correlation between $k$ and $M_{\rm BH}(t_{23})/M_*(t_{23})$ is 
seen from the scattered points, and confirmed by the negative (rank-based) 
Spearman correlation coefficient.
The dashed line in Fig.~\ref{fig:phase-3-self-regularization} shows a linear
regression of $k$ on $\log\,\left[M_{\rm BH}(t_{23})/M_*(t_{23})\right]$, and has a  
slope of $\approx -3.0$. 
This slope indicates that, if a SMBH deviates from the equilibrium scaling
relation (Eq.~\ref{eq:phase-3-scaling}), it is pulled back
once $M_{*}$ grows for about $0.3\dex$. For a typical main-sequence 
galaxy at $z \approx 0$ with sSFR $\sim 1\Gyr^{-1}$, this pull-back 
timescale is $\sim 2 \Gyr$, shorter than the Phase-3 duration of most 
galaxies (Fig.~\ref{fig:fit-info}). Hence, most SMBHs are well 
self-regulated during Phase-3, which erases the footprints of their early 
growth, including the seeding process. 
{\revised We have performed additional tests using star-forming galaxies 
at $z=0$ whose Phase-3 is well established, and found that the anti-correlation 
between $k$ and $M_{\rm BH}(t_{23})/M_*(t_{23})$ is still present, not 
limited by our sample selection (see \S\ref{sec:data-and-sample}).
The self-regulation dominated growth also} 
explains the tight relation between $M_{\rm BH}$ and $M_*$ 
at $z=0$, as shown by the black solid curve in Fig.~\ref{fig:fit-info}.
Observations also suggest a relation between $M_{\rm BH}$ and $M_*$ 
(or $M_{\rm *,bulge}$) in the local Universe 
\citep[e.g.][]{Kormendy2013,greeneIntermediateMass2020,
zhang2024HaloMassobservableProxy},
which we will discuss in \S\ref{ssec:obs-scaling}.

The similar trend of sHAR, sSFR, and sBHAR during 
Phase-3 shown in the top left panel of Fig.~\ref{fig:phase-2-mass-history} 
suggests that the rate of evolution of a
galaxy in the $M_{\rm BH}$-$M_{\rm *}$ plane is controlled by 
the accretion rate of the host halo. 
The regulation by the AGN feedback bridges the growth of
SMBH with its environment, transfers the growth pattern of the host halo 
to the galaxy and SMBH, and leads to the establishment of the scaling
relation derived in Eq.~\eqref{eq:phase-3-scaling}.
The sBHAR decreases with time during Phase-3, as it follows the sHAR that in general 
decreases with time. Once a certain mass threshold is reached, the mode of AGN 
feedback implemented in TNG makes a transition, as we will discuss in the following sections.

\subsubsection[]{The trigger of the kinetic-mode feedback} 
\label{sssec:phase-3-triger-kin}

\begin{figure*}[htb]
\centering
\includegraphics[width=.9\textwidth]{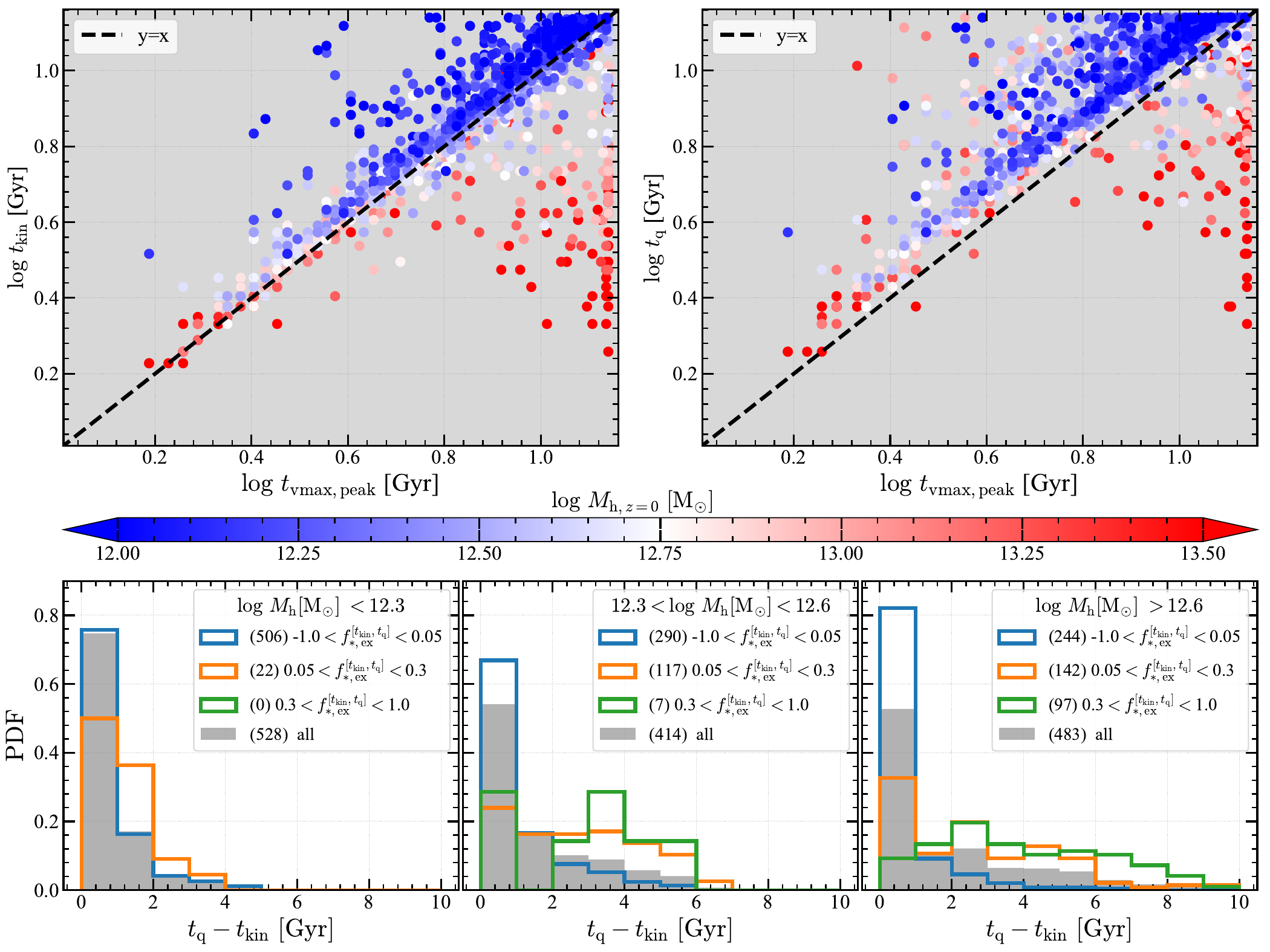}
\caption{
    Distribution of the properties relevant to the quenching of galaxies.
    {\bf Top panels} show the distributions of individual galaxies in the 
    $t_{\rm kin}$-$t_{\rm vmax,peak}$ ({\bf left}) and 
    $t_{\rm q}$-$t_{\rm vmax,\,peak}$ {\revised 
    ({\bf right})} planes, color-coded 
    by their $M_{\rm h}$ at $z=0$.
    {\bf Bottom panels} show the histograms of $t_{\rm q}-t_{\rm kin}$, the 
    time lag of quenching after the trigger of the kinetic-mode feedback.
    The {\bf grey} histogram in each panel shows the result
    for a subsample of galaxies in a given range of $M_{\rm h}$ at $z=0$.
    This subsample is further divided into three bins by 
    $f^{[t_{\rm kin},t_{\rm q}]}_{\rm *, ex}$, the ex-situ growth of stellar 
    mass between $t_{\rm kin}$ and $t_{\rm q}$, and their histograms 
    are shown in different colors.
    The number of galaxies in each bin is labeled in the legend.
    See \S\ref{sssec:phase-3-triger-kin} and \S\ref{sssec:quenching-criterion} 
    for the definitions of these times and their implications for galaxy 
    quenching.
}
\label{fig:phase-3-time-sequence} 
\end{figure*}

TNG implements a two-mode AGN feedback: a thermal mode
at high Eddington ratios and a kinetic mode at low Eddington ratios.
The kinetic mode is implemented to effectively transport the 
feedback energy to the gas at a large scale and to prevent over-cooling of 
the halo gas \citep[e.g.][]{weinbergerSimulatingGalaxyFormation2017,
weinbergerSupermassiveBlackHoles2018,pillepichSimulatingGalaxyFormation2018a}.
Thus, the transition of AGN feedback from the thermal to the kinetic mode
modifies the growth of the galaxy and the SMBH at late times, 
and is critical in shaping the observational properties, in particular those 
related to quenching, of low-$z$ galaxies predicted by the simulation.
Here, we search for a criterion for this transition in order to understand 
how it can relate to the properties of the galaxies and halos at low $z$.

In TNG, the kinetic-mode feedback is triggered when the Eddington ratio, 
$\chi = \dot{M}_{\rm BH}/\dot{M}_{\rm Edd}$, falls below a threshold
\begin{equation}
    \chi_{\rm thres} \equiv \min \left[ 
        0.002 \left(\frac{M_{\rm BH}}{10^8 \Msun}\right)^2, 0.1 \right] \,,
\end{equation}
where the $M_{\rm BH}$-dependence is introduced to favor the transition 
of SMBHs in massive galaxies so that their quenched fraction is consistent
with observations, and the upper bound, $0.1$, is set to avoid 
a too-early stop of the SMBH growth in massive galaxies
\citep[see \S2.1 of][]{weinbergerSimulatingGalaxyFormation2017}. 
By definition, the Eddington limit $\dot{M}_{\rm edd}$ is proportional to 
$M_{\rm BH}$. 
If the gas environment is only slowly evolving during Phase-3 (see, e.g.
the slow evolution of $A_{\rm s}$ in the top right panel of 
Fig.~\ref{fig:phase-2-mass-history}), 
$\chi$ is proportional to the sBHAR. According to the self-regulation 
scenario described in 
\S\ref{ssec:smbh-self-regularization}, the evolution of sBHAR follows that of 
sHAR (see the top left panel of Fig.~\ref{fig:phase-2-mass-history}). Thus,
$\chi$ is also roughly proportional to the sHAR.

For low-mass SMBHs {\revised(a subsample with $M_{{\rm BH},z=0} \lesssim 10^{8} (2 \times 10^{8}) \Msun$, 
containing 38(578) galaxies, about 2.3 (40.5)\% of our total sample, and with a 
median halo mass of $M_{{\rm h},z=0} = 10^{11.8} (10^{12.2}) {\rm M}_{\odot}$}), the threshold 
$\chi_{\rm thres}$ is close to zero, and thus the transition occurs only 
if the sHAR approaches zero. 
A universal feature of halo assembly is the presence of two phases: 
a fast-accretion phase at high redshift, when the halo potential rapidly deepens, 
and a slow-accretion phase at low redshift, when the halo potential remains
constant or even declines \citep[e.g.][]{zhaoGrowth2003a}. 
Thus, for the SMBH in a low-mass halo to trigger the kinetic-mode feedback, its 
host halo must transit to the slow phase. 
To verify that the halo transition indeed leads to the trigger of the kinetic-mode 
feedback, we follow \citet{zhaoGrowth2003a} to define the halo transition time 
as $t_{\rm vmax,\,peak}$, the time when the maximum of the circular 
velocity, $V_{\rm max}$, reaches the peak,
{\revised and we follow \citet[see their \S3.1]{weinbergerSupermassiveBlackHoles2018}
to define whether the AGN feedback is in the thermal or kinetic mode.
The last time when the AGN feedback switches from the thermal to kinetic mode
is found, and denoted as $t_{\rm kin}$, for each galaxy in our sample.
Only two galaxies in our sample are quenched at $z = 0$ without activating 
the kinetic-mode feedback, and we remove them from 
the analyses involving $t_{\rm kin}$.
We compare $t_{\rm kin}$ with $t_{\rm vmax,\,peak}$ for individual halos,
and show the results in the top left panel of}
Fig.~\ref{fig:phase-3-time-sequence}, color-coded according 
to their halo masses at $z=0$.
For halos with $M_{{\rm h}, z=0} \lesssim 10^{12.5} \Msun$, $t_{\rm vmax,\,peak}$
correlates well with $t_{\rm kin}$, indicating that the transition of
the AGN feedback is indeed driven by the transition of halo accretion.

For massive SMBHs (a subsample with $M_{{\rm BH}, z=0} \gtrsim 7 \times 10^{8} \Msun$,
{\revised containing 155 galaxies, about 10.9\% of our total sample, 
and with a median halo mass of $M_{{\rm h}, z=0} = 10^{13.4} {\rm M}_{\odot}$}), 
the threshold $\chi_{\rm thres}$ approaches 
the upper limit $0.1$. The transition of these halos can occur even 
when they are fast accreting. To derive the condition
for their transition, we follow 
\citet[see their \S5.1]{dekelEfficientFormationMassive2023}
to approximate the specific accretion rate of a fast-accreting halo as
\begin{equation}
    \dot{M}_{\rm h} / M_{\rm h} \propto M_{\rm h}^{0.14} (1+z)^{5/2}\,.
\end{equation}
This rate depends only weakly on $M_{\rm h}$ but strongly on redshift, 
suggesting that the Eddington ratio of a self-regulated SMBH 
in a massive halo depends only on redshift. 
A detailed study by \citet[see their figure 6]{weinbergerSimulatingGalaxyFormation2017}
indeed shows such a strong redshift dependence and weak halo-mass dependence.
Given this property of the specific growth rate of halo mass, 
the condition for the transition of AGN feedback within massive halos 
becomes 
\begin{equation}
    \chi \sim (1+z)^{5/2} \sim \chi_{\rm thres} = 0.1\,,
\end{equation}
which depends only on redshift and indicates that SMBHs in massive halos
transit roughly at a fixed redshift. The red dots in the top left panel of Fig.~\ref{fig:phase-3-time-sequence} 
for halos with $M_{\rm h} \gtrsim 10^{13} \Msun$ show that $t_{\rm vmax,peak}$
can span a wide range, and many of such halos continue to accrete
fast until $z=0$. On the other hand, $t_{\rm kin}$ can be much earlier 
than $t_{\rm vmax,peak}$, with most of the halos located around
$t_{\rm kin} =3\Gyr$ ($z \approx 2$), which confirms our argument.

The SMBHs in intermediate-mass halos fall between the two extreme cases
and the trigger of the kinetic mode is at a time between $3 \Gyr$ and $t_{\rm vmax,\, peak}$, 
as shown by the white dots in the top left panel of Fig.~\ref{fig:phase-3-time-sequence}. 
We note that this conclusion is a result of the specific choice of $\chi_{\rm thres}$ in TNG. 
If, for example, the threshold $\chi_{\rm thres}$ is set to be a constant of 0.1 for all halos, 
the transition of the AGN feedback would be earlier for low-mass halos.

\subsubsection{The quenching of galaxies}
\label{sssec:quenching-criterion}

\begin{figure*}
    \centering
    \includegraphics[width=.9\textwidth]{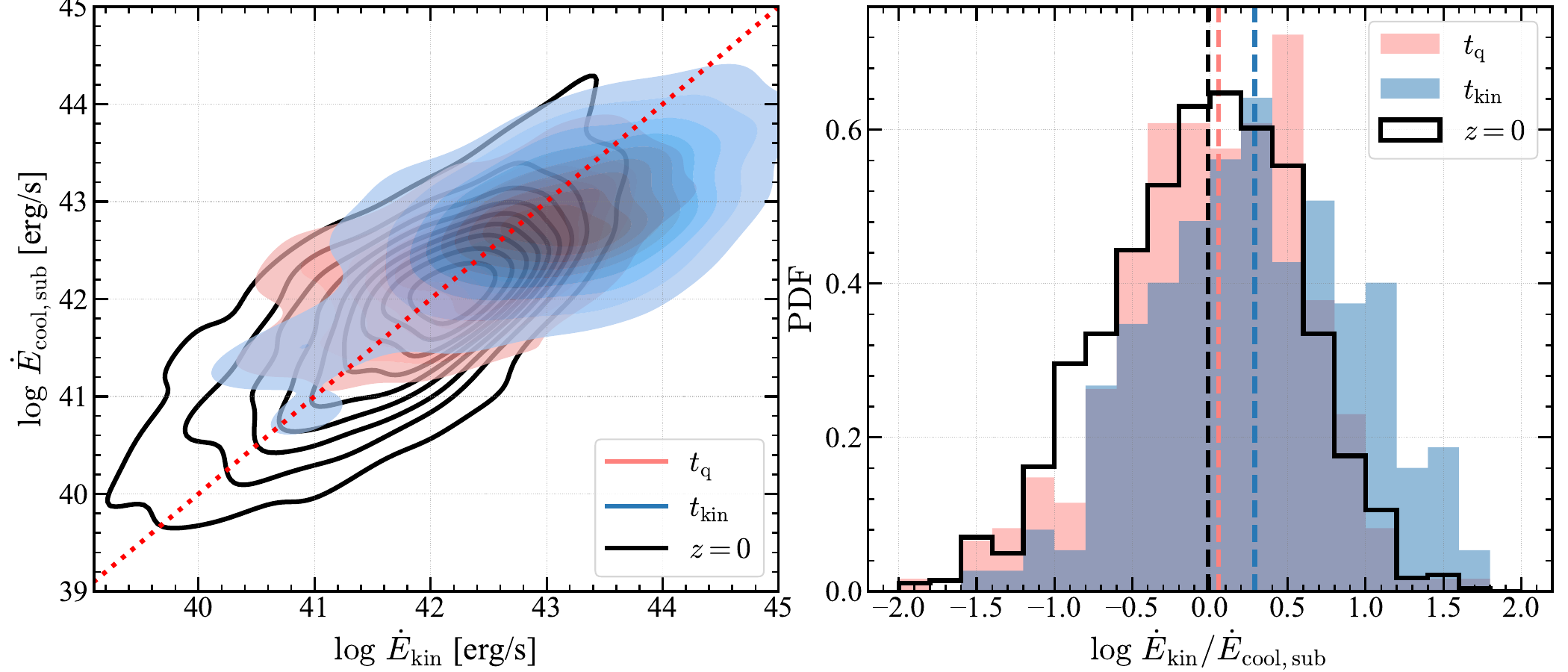}
    \caption{
        The sufficient condition for the quenching of galaxies.
        {\bf Left panel} shows the distribution of galaxies in the 
        $\dot{E}_{\rm cool,sub}$-$\dot{E}_{\rm kin}$ plane, where $\dot{E}_{\rm cool,sub}$ is
        the cooling rate of all non-star-forming gas cells in the subhalo,
        and $\dot{E}_{\rm kin}$ is the energy output rate from kinetic-mode AGN.
        {\bf Contours} with a given color, {\revised 
        from inner to outer, encompass 10\% -- 95\% of the galaxies
        in our sample at a given time point ($t_{\rm q}$, $t_{\rm kin}$ or $z=0$) 
        during their evolution histories along the main branch}. 
        Limited by the data storage, cooling rates are only available in
        the 20 ``full snapshots'' of TNG. {\revised 
        Thus, the contour for 
        $t_{\rm kin}$ (or $t_{\rm q}$) only includes the galaxies whose 
        $t_{\rm kin}$ ($t_{\rm q}$) is in the ``full snapshots'',
        which reduces the sample size to 199 (304). }
        {\bf Dashed line} indicates where these two rates are equal.
        {\bf Right panel} shows the histograms of the ratio, 
        $\dot{E}_{\rm kin}/\dot{E}_{\rm cool,sub}$, at different time points.
        {\bf Dashed lines} indicate the medians of the distributions.
        See \S\ref{sssec:quenching-criterion} for the details.
    }
    \label{phase-3-energy-balance} 
\end{figure*}

The feedback energy in the kinetic mode implemented by TNG
affects the gas effectively by both ejection and prevention, and can  
quickly reduce the amount of star-forming gas in the galaxy and beyond 
\citep[e.g.][]{zingerEjectivePreventativeIllustrisTNG2020,
shi2022ColdGasMassive,
martin-navarro2021AnisotropicSatelliteGalaxy}.
Consequently, galaxies are expected to be quenched quickly after the transition 
of the feedback mode. In the top right panel of Fig.~\ref{fig:phase-3-time-sequence}, 
we show $t_{\rm q}$, 
the time when a galaxy is quenched {\revised (see \S\ref{sec:data-and-sample} for 
the definition)}, and compare it to $t_{\rm vmax,\,peak}$ 
for individual galaxies. 
For all galaxies,  $t_{\rm q}$ is systematically
later (larger) than $t_{\rm kin}$, indicating that the kinetic-mode AGN feedback 
is a necessary condition for quenching central galaxies, and that 
galaxies may sustain their star formation for some time after the trigger of 
the kinetic-mode feedback.

To quantify the time needed for the kinetic-mode feedback to quench 
galaxies, the grey histograms in the three bottom panels of 
Fig.~\ref{fig:phase-3-time-sequence}
show the distributions of $t_{\rm q} - t_{\rm kin}$ for halos with different 
masses at $z=0$. For the least massive halos ($M_{{\rm h}, z=0} < 10^{12.3} \Msun $),
the majority of the probability mass is located within $1\Gyr$, indicating
that their quenching is very rapid.
For more massive halos ($M_{{\rm h}, z=0} > 10^{12.3}\Msun$),  
a long tail emerges in the distribution, indicating that  
a noticeable fraction of galaxies remain star-forming for more than 
$4\Gyr$ after the trigger of the kinetic-mode feedback.
The increase of the time lag of quenching, $t_{\rm q} - t_{\rm kin}$,
with increasing halo mass indicates that central galaxies
residing in massive halos can acquire star-forming gas via 
additional channels. As suggested by \citet{moTwophaseModelGalaxy2024},
a more massive halo remains in the phase of fast accretion
down to a lower redshift than a less massive one,
and thus is expected to bring in more satellite galaxies that can eventually 
merge with the central galaxy. 
As suggested by observations \citep[e.g.][]{dahlem1994StarFormationTriggered,
sharmaIkaModelFeedbackregulated2020} and hydrodynamical simulations 
\citep[e.g.][]{capeloShocksAngularMomentum2017,peschken2020DiscGalaxiesFormed,
karam2024DynamicsStarCluster}, mergers 
can effectively redistribute angular momentum and induce gas inflow, and also 
compress the gas to trigger rapid cooling and star formation. 
The longer time lag of quenching in more massive halos is thus a natural 
consequence of their higher merger rates. To verify this expectation,
we divide the sample of galaxies in each bottom panel of Fig.~\ref{fig:phase-3-time-sequence} 
into subsamples according to the ``ex-situ'' growth of stellar mass,
\begin{equation} \label{eq:frac-star-ex-situ}
    f_{\rm *,ex}^{[t_{\rm kin}, t_{\rm q}]} \equiv
    \Delta M_{\rm *,ex}^{[t_{\rm kin}, t_{\rm q}]} /M_*(t_{\rm q})   \,,
\end{equation}
where $\Delta M_{\rm *,ex}^{[t_{\rm kin}, t_{\rm q}]}$ is the mass of stars 
formed ex-situ during the time interval between $t_{\rm kin}$ and $t_{\rm q}$
and fallen into the galaxy before $t_{\rm q}$ via either stripping or merger,
and $M_*(t_{\rm q})$ is the stellar mass of the galaxy at $t_{\rm q}$. 
{\revised We utilize the published catalog generated by 
\citet{rodriguez-gomez2015MergerRateGalaxies} and 
\citet{rodriguez-gomezStellarMassAssembly2016a}
to identify the ex-situ stars for each galaxy, defined as
those bound to the galaxy but formed outside the main branch of the subhalo 
merger tree rooted in the galaxy.}
The distribution of the time lag of quenching for each subsample is 
shown in the bottom panels of Fig.~\ref{fig:phase-3-time-sequence}
by a colored histogram.
For galaxies with $f_{\rm *,ex}^{[t_{\rm kin}, t_{\rm q}]} < 5\%$, 
the time lag is sharply peaked within $1\Gyr$, regardless of the halo mass. 
This demonstrates that the kinetic-mode feedback is indeed effective 
in quenching the galaxy in the absence of merger-induced gas inflow.
However, for galaxies with strong mergers ($f_{\rm *, ex}^{[t_{\rm kin}, t_{\rm q}]} > 30\%$), 
star formation can sustain a long time, even if the kinetic-mode feedback
is triggered. The quenching of a galaxy is thus delayed to 
a time when the frequency of mergers falls below a certain level, 
depending on the sSFR threshold used to define the quenching
(\S\ref{sec:data-and-sample}).
A similar conclusion was reached by \citet{terrazasRelationship2020}, where 
the cumulative energy of AGN is computed and compared to the binding energy
of the gas in a galaxy. Galaxies are found to be quenched only when 
the feedback energy exceeds about 100 times the binding energy of the gas,
indicating that most of the feedback energy is lost due to radiative cooling.

The cease of the merger marks the end of the stochastic creation of star-forming
gas, leaving the cooling of hotter gas in a halo being the only channel for the central
galaxy to maintain its star formation. In this case, the gas surrounding the SMBH is dominated 
by that in the hot phase, as shown in the bottom right panel of Fig.~\ref{fig:phase-2-mass-history}.
If the energy injection by the kinetic-mode feedback remains strong
even when the amount of gas available for the SMBH to accrete is small,      
as is the case in TNG because of its large feedback efficiency at low accretion rates 
\citep[][see their \S2.3]{weinbergerSimulatingGalaxyFormation2017},
the amount of the gas within the galaxy will continue to decrease due to both 
ejection and prevention, causing the rates of gas cooling and star formation
to slow down. As shown in the top-right panel of Fig.~\ref{fig:phase-2-mass-history}, 
the effective sound speed of the gas during Phase-3 continues to increase, while the 
density decreases, both leading to a decrease in $A_{\rm s}$ and thus 
to the growth of the SMBH and the strength of its feedback.
Such an evolution stops once the feedback of AGN is reduced to a level
comparable to the cooling rate of the gas. This final stage, where 
cooling, star formation, SMBH growth, and AGN feedback are all slow,
produces a quenched state that is consistent with observations.

{\revised
The above argument leads to an equality that characterizes the condition 
for the kinetic-mode AGN feedback to maintain the quenching of a galaxy
without significant mergers, which is that the feedback balances the 
cooling of gas:}
\begin{equation} \label{eq:sufficient-condition-quenching}
    \dot{E}_{\rm cool, sub} = \dot{E}_{\rm kin}\,,
\end{equation}
where $\dot{E}_{\rm cool, sub}$ is the total cooling rate of all 
non-star-forming gas cells within the host subhalo of the galaxy,
$\dot{E}_{\rm kin} \equiv \eta_{\rm kin} \dot{M}_{\rm BH} c^{2}$ 
is the energy output rate of the kinetic-mode AGN, and  
$\eta_{\rm kin} = {\rm min}\left[ \rho / (0.05\rho_{\rm SF,\,thresh}),\, 0.2 \right]$ is 
a coupling efficiency \citep[][see their Eq.~5]{weinbergerSupermassiveBlackHoles2018}.
The inclusion of the subhalo gas outside the galaxy is necessary
for TNG, as its kinetic-mode feedback works through both ejection and prevention
over a large scale.
The contours in the left panel of Fig.~\ref{phase-3-energy-balance} show 
the distribution of galaxies in the $\dot{E}_{\rm cool, sub}$-$\dot{E}_{\rm kin}$
plane at different epochs, while the histograms in the right panel
show the marginal distributions of $\dot{E}_{\rm kin}/\dot{E}_{\rm cool, sub}$,
both in logarithmic scales.
As expected, at the trigger time of the kinetic-mode feedback, $t = t_{\rm kin}$, 
both $\dot{E}_{\rm kin}$ and $\dot{E}_{\rm cool, sub}$ are systematically 
larger than those at later times. In particular,
$\dot{E}_{\rm kin}$ is larger than $\dot{E}_{\rm cool, sub}$, 
indicating a non-equilibrium and the continuing reduction of gas.
At the quenching point, $t=t_{\rm q}$, the two rates become balanced,
both significantly smaller than those at $t_{\rm kin}$, implying 
a stable state of quenching and the transition of the galaxy into 
the final phase.

\subsection{Phase-4: Merger dominated growth} \label{ssec:phase-4}

\begin{figure*}[htb]
\centering
\includegraphics[width=.9\textwidth]{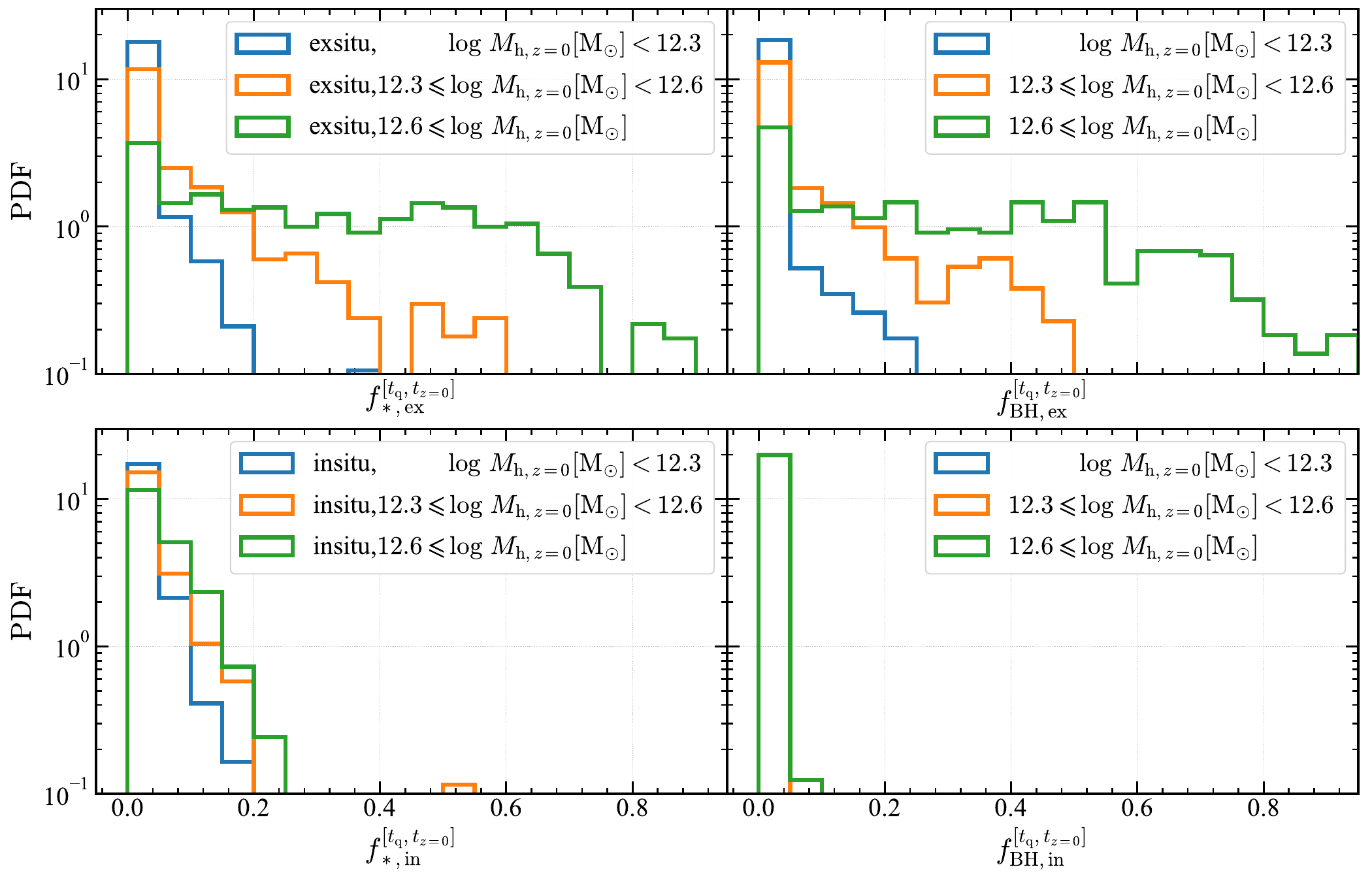}
\caption{
    Distributions of the ex-situ ({\bf upper row}) and in-situ 
    ({\bf bottom row}) fractions of $M_*$ ({\bf left column}) and 
    $M_{\rm BH}$ growth ({\bf right column}), 
    all evaluated within the time interval between $t_{\rm q}$ and $z=0$.
    In each panel, histograms with different colors show the distributions
    of galaxies with different $M_{\rm h}$ at $z=0$.
    This figure suggests that the growth of galaxies and SMBHs after quenching 
    (during Phase-4) is dominated by mergers, especially for massive halos.
    See \S\ref{ssec:phase-4} for the details.
}
\label{fig:phase-4-merge} 
\end{figure*}

The final stage of the evolution is a quenched phase.
In this phase, the in-situ growth of SMBHs and galaxies ceases, while the 
ex-situ growth driven by galaxy-galaxy and SMBH-SMBH mergers continues. 
In the top left panel of Fig.~\ref{fig:phase-4-merge}, 
we show the distribution of $f_{\rm *,ex}^{[t_{\rm q}, t_{z=0}]}$,
the fraction of ex-situ growth of stellar mass between $t_{\rm q}$ and $z=0$, 
defined similarly to Eq.~\eqref{eq:frac-star-ex-situ} but with a 
different time interval, for halos 
with different $M_{{\rm h},z=0}$. For comparison, in the other three 
panels, we show the distributions of the fraction of in-situ growth of
stellar mass (bottom left), and the ex- and in-situ growth of SMBH mass
(right column), respectively.
For the ex-situ growth of stellar mass, a clear trend with halo mass
is seen, with more massive halos having a larger fraction of ex-situ growth
after quenching. 
This is expected as a consequence of the larger number of satellite galaxies
in more massive halos \citep[e.g.][]{yang2009GALAXYGROUPSSDSS,
yang2012EVOLUTIONGALAXYDARK,wang2024MeasuringConditionalLuminosity}
carried in by halo-halo mergers.
For halos with $M_{{\rm h},z=0} \geqslant 10^{12.6} \Msun$, there are 
cases where the ex-situ fraction exceeds $80\%$, indicating that 
the final properties of the central galaxies in these halos are shaped by 
their merger histories. These merger-dominated galaxies are expected
to form in massive halos with late formation times, as shown by 
\citet{chen2023MassiveDarkMatter} using an N-body simulation constrained by 
the density field that matches the observed structures 
found in SDSS \citep{wangELUCIDEXPLORINGLOCAL2016}. A plausible example 
is the Coma cluster, whose central subhalo appears at $z \approx 5$, 
a relatively late time in comparison with other massive halos,
and then grows rapidly via a large number of major mergers. The BCG of Coma is thus 
likely to reside in the tail of the distribution of 
$f_{\rm *,ex}^{[t_{\rm q}, t_{z=0}]}$.
On the other hand, the fraction of in-situ growth of stellar mass
since $t_{\rm q}$ is nearly independent of halo mass, and nearly always 
below $20\%$. 
Earlier works have suggested that significant in-situ growth of stellar mass
can be induced by mergers, via the perturbation of angular momentum and 
the compression of gas by shocks \citep[e.g.][]{capeloShocksAngularMomentum2017,lagosQuantifyingImpactMergers2018}.
Our result here suggests that, in TNG, the in-situ growth of stellar mass in 
galaxies that have already become ``dry'' (gas-poor) is insignificant.

The distribution of in- and ex-situ fractions of $M_{\rm BH}$ growth is similar 
to those of $M_*$. Specifically, the ex-situ fraction of $M_{\rm BH}$ growth
increases with $M_{{\rm h},z=0}$, and in extreme cases, can exceed $80\%$.
The in-situ fraction of $M_{\rm BH}$ growth is always small, 
nearly zero, due to the strong kinetic-mode feedback that clears the gas
surrounding the SMBH. The result here is consistent with 
\citet[][see their \S4.1 and Fig.7]{weinbergerSupermassiveBlackHoles2018},
where the growth of SMBHs in massive halos experiences a transition, 
from being dominated by in-situ accretion during the phase of thermal-mode 
feedback to being dominated by mergers during the phase of kinetic-mode 
feedback. As mergers tend to bring galaxies into a 1:1 relation in the 
$M_{\rm BH}$-$M_*$ plane, the $M_{\rm BH}$-$M_*$ scaling relation is 
expected to be modified during Phase-4. We will discuss this in 
\S\ref{ssec:obs-scaling}.

As the kinetic-mode AGN feedback in TNG is designed to sustain the quenching 
of massive galaxies via the self-regulated accretion 
\citep{weinbergerSimulatingGalaxyFormation2017}, the sufficient condition 
(Eq.~\ref{eq:sufficient-condition-quenching}) 
of feedback-cooling balance for quenching should remain valid
over the entire phase of the quenched evolution. 
The black contour in Fig.~\ref{phase-3-energy-balance} shows 
a comparison of the cooling rate within the subhalo and the 
feedback energy rate at $z=0$. 
Both rates are shifted to lower values relative to those at $t_{\rm q}$, 
but remain around the balance line. 
Indeed, observations have suggested that the atmosphere of massive galaxies 
is in a state of marginally stable to form cold clouds, and the baryon cycling induced by 
self-regulating AGN feedback is capable of explaining the observed
quasi-equilibrium state of massive galaxies \citep[see, e.g.][for a review]{donahue2022BaryonCyclesBiggest}.
A similar conclusion was reached by \citet{zingerEjectivePreventativeIllustrisTNG2020} using TNG, 
where gas entropy is used to quantify the effect of AGN feedback. 
Since high-entropy gas 
can have high temperature and thus high specific internal energy to be dispersed 
before forming stars, and have low cooling rate due to the high resistance to 
compression, the use of entropy as a proxy for the feedback effect is intuitive.
Their results show that the entropy of the gas in the circumgalactic medium (CGM) of 
massive galaxies is significantly lifted by the kinetic-mode feedback out to a distance of
several hundred kiloparsecs, so that quenching can be sustained.
{\revised An interesting side effect of the large-scale  
kinetic-mode feedback, as proposed by
\citet{martin-navarro2021AnisotropicSatelliteGalaxy},
is the reduction of ram pressure stripping of satellite galaxies along the
minor axes of the central galaxies where the gas is preferentially
cleared by the feedback from the central galaxies. 
However, as clearly demonstrated by 
\cite{kangAlignmentSatellitesCentral2007} and \cite{karpAnisotropicSatelliteGalaxy2023}, 
the angular variation of subhalo properties in halos also produces 
significant anisotropy in the properties of satellite galaxies 
around central galaxies. A careful subtraction of the effects of
subhalo anisotropy is thus necessary to isolate the effect of the
AGN feedback on satellite galaxies at large scales.}

\section{Reconstructing the phases of SMBH growth from observations} 
\label{sec:observation}

\subsection{The build-up of scaling relations}
\label{ssec:obs-scaling}

\begin{figure*}[htb]
\centering
\includegraphics[width=.9\textwidth]{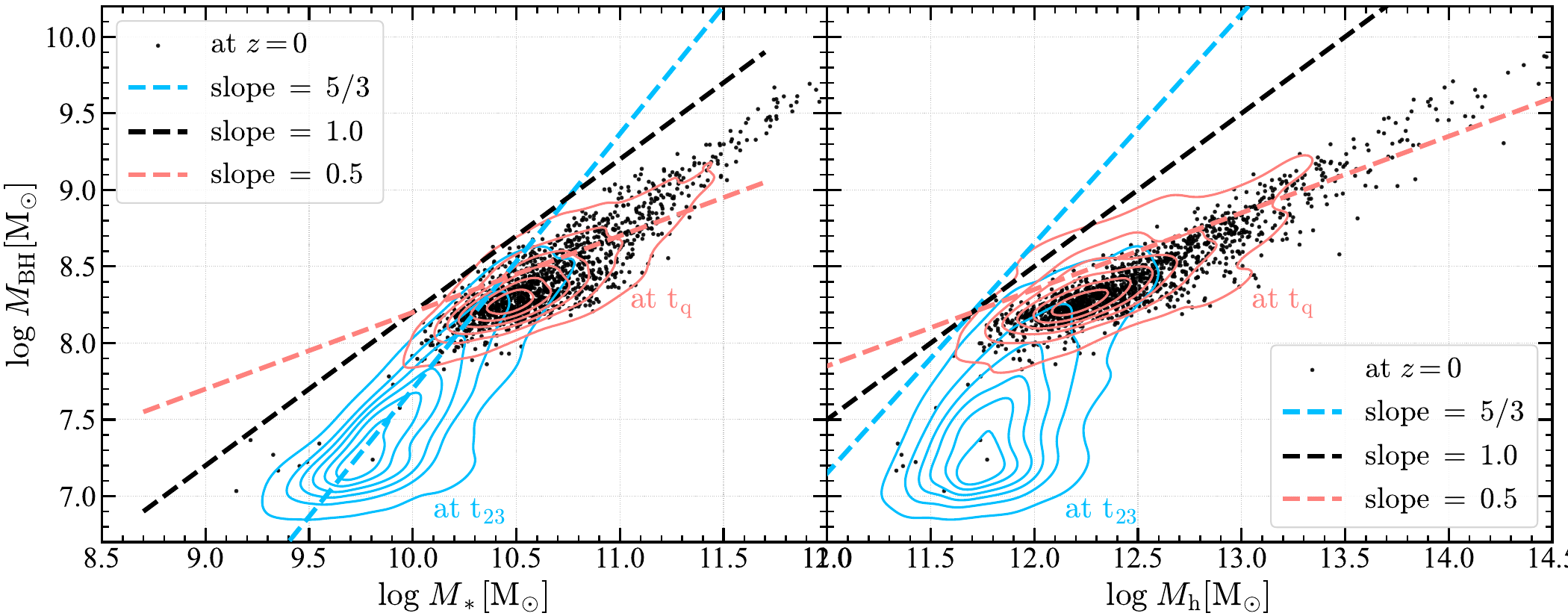}
\caption{
    $M_{\rm BH}$-$M_*$ ({\bf left panel}) and 
    $M_{\rm BH}$-$M_{\rm h}$ ({\bf right panel})
    relations at different time points, 
    $t_{23}$ ({\bf blue contours}), $t_{\rm q}$ ({\bf red contours}), 
    and $z=0$ ({\bf black dots}), in the histories of the galaxies.
    In each panel, three dashed lines indicate three relations 
    with different logarithmic slopes.
    Contours from inner to outer enclose 10\% -- 95\% of the sample.
    See \S\ref{ssec:obs-scaling} for the details about the build-up of these 
    relations.
}
\label{fig:scaling-init} 
\end{figure*}

As discussed in \S\ref{sec:processes}, the evolution of a SMBH (and its
host galaxy/halo) involves a series of critical times that 
mark the transitions under specific conditions.
In the $M_{\rm BH}$-$M_*$ (or $M_{\rm BH}$-$M_{\rm h}$) plane, 
these conditions form a series of edges that a SMBH has to pass before 
it can reach subsequent phases. 
SMBHs with different initial conditions start at different positions
in the plane, ``flow'' in the plane following different paths,
and cross the edges at different locations and times.
Thus, the $M_{\rm BH}$-$M_*$ (or $M_{\rm BH}$-$M_{\rm h}$) scaling relations 
at a given epoch for a population of SMBHs in a given phase 
can be viewed as a time slice of the flow lines in the plane 
between the two edges that bracket the phase in question.
If observations are able to obtain such slices of different redshifts,
the locations of the edges can be inferred, thus providing constraints
on the physical processes that drive the transitions in the evolution.

Fig.~\ref{fig:scaling-init} shows the distributions of SMBHs in the 
$M_{\rm BH}$-$M_*$ and $M_{\rm BH}$-$M_{\rm h}$ planes at three times: 
$t_{23}$, when the fast accretion of SMBH ends,
$t_{\rm q}$ ($ \equiv t_{34}$), when the galaxy is quenched, 
and $z=0$. As discussed in \S\ref{ssec:phase-2}, the transition at $t_{23}$
happens when the thermal-mode feedback of AGN is able to unbind the gas 
of the galaxy. According to Eqs.~\eqref{eq:condition-t-23}
and \eqref{eq:phase-3-scaling}, this corresponds to a condition of 
\begin{equation} \label{eq:scaling-t-23}
    \eta_{\rm eff} \eta_{\rm thermal} M_{\rm BH} 
    \gtrsim f_{\rm gas} M_{\rm h}^{5/3} 
    = f_{\rm gas} f_*^{-5/3}M_*^{5/3} \,,
\end{equation}
which predicts a power-law relation between $M_{\rm BH}$ and $M_*$ (or $M_{\rm h}$),
with a power index $\gamma = 5/3$ at the time when  
the star formation efficiency peaks, or a different index if 
$f_{\rm gas}$ and $f_*$ are mass-dependent. 
As shown by the blue contours in the left panel of Fig.~\ref{fig:scaling-init},
the $M_{\rm BH}$-$M_*$ relation at $t_{23}$ follows  
a power-law with a power index of about $5/3$. This implies a nearly constant 
$f_{\rm gas} f_*^{-5/3}$ at around $t_{\rm 23}$, and suggests that 
the gas and stellar masses are positively correlated.
At $M_*\approx 10^{10} \Msun$ and $M_{\rm BH} \approx 10^7 \Msun$, there is a 
population of SMBHs that deviates from the relation. As noted in \S\ref{sec:def-phases}, 
this is mainly due to the difficulty in determining $t_{23}$ accurately.
On the other hand, as shown by the blue contours in the right
panel, the $M_{\rm BH}$-$M_{\rm h}$ relation in logarithmic scales appears to be 
non-linear even for SMBHs with $M_{\rm BH}$ well above $10^7 \Msun$. This reflects
the non-linear dependence of $f_{\rm gas}$ on $M_{\rm h}$. At 
about $M_{\rm h} = M_{\rm h,c} = 10^{12} \Msun$, $f_{\rm gas}$ reaches 
its peak, leading to a $M_{\rm BH}$-$M_{\rm h}$ relation that follows
the power-law of $\gamma=5/3$. At lower or higher $M_{\rm h}$, the
dependence of $f_{\rm gas}$ on $M_h$ changes, which results in bending
in the $M_{\rm BH}$-$M_{\rm h}$ relation. 

As concluded in \S\ref{ssec:phase-3}, the evolution of SMBHs between  
$t_{23}$ and $t_{\rm q}$ is self-regulated so as to follow $M_{\rm BH} \propto M_*^{5/3}$.
Naively one would expect that the $M_{\rm BH}$-$M_*$
relation at $t_{\rm q}$ follows the same power law.
However, as shown by the red contours in the left panel of 
Fig.~\ref{fig:scaling-init}, the $M_{\rm BH}$-$M_*$ relation at $t\sim t_{\rm q}$
appears to be much shallower, with $\gamma<1$.
As discussed in \S\ref{sssec:phase-3-triger-kin}, the trigger of the kinetic-mode
feedback is controlled by the $M_{\rm BH}$-dependent threshold of the 
Eddington ratio, $\chi_{\rm thres}$, for SMBHs with 
$M_{\rm BH} \lesssim 7 \times 10^8 \Msun$.
It can also be seen from Fig.~\ref{fig:phase-3-time-sequence} that the majority 
of galaxies are quenched soon (within $\approx 1\Gyr$) after the trigger of the 
kinetic-mode feedback for halos with $M_{\rm h} < 10^{12.3} \Msun$.
In TNG, $\chi_{\rm thres}$ is set so that the kinetic-mode feedback 
is triggered preferentially for high-mass SMBHs. 
Thus, 
the quenching of high-mass galaxies occurs earlier than that 
expected from a model with a constant (and small) $\chi_{\rm thres}$.
The shallow $M_{\rm BH}$-$M_*$ relation at $M_{\rm *} \lesssim 10^{11} \Msun$ 
is a direct consequence of the choice of the $\chi_{\rm thres}$-$M_{\rm BH}$ 
relation. For higher masses ($M_{\rm BH} \gtrsim 7 \times 10^8 \Msun$), 
the threshold $\chi_{\rm thres}$ reaches the upper 
limit, 0.1, and becomes independent of $M_{\rm BH}$. Meanwhile, the 
conditions necessary for the quenching of galaxies in high-mass halos
are delayed by frequent mergers (\S\ref{sssec:quenching-criterion}). 
Eventually, the preference for quenching in high-mass galaxies is 
weakened, which allows the SMBHs in them to grow further. The tip present in
the red contours at $M_* \gtrsim 10^{11}\Msun$ in the
left panel of Fig.~\ref{fig:scaling-init} verifies this.
The $M_{\rm BH}$-$M_{\rm h}$ relation at $t_{\rm q}$, as shown 
in the right panel, appears as shallow as $\gamma=0.5$. 
The difference in the index between $M_{\rm BH}$-$M_*$ and $M_{\rm BH}$-$M_{\rm h}$
relations at $t_{\rm q}$ can be understood by an argument similar to that 
used for the bending $M_{\rm BH}$-$M_{\rm h}$ relation at $t_{\rm 23}$, but here
both $f_{\rm gas}$ and $f_*$ decrease with increasing $M_{\rm h}$ 
at $M_{\rm h} \gtrsim 10^{12} \Msun$.

The evolution of SMBHs (and their host galaxies) in the final phase 
after $t_{\rm q}$ is dominated 
by the ex-situ growth. As shown in \S\ref{ssec:phase-4}, the fraction of 
ex-situ growth for both $M_{\rm BH}$ and $M_*$ depends significantly 
on the halo mass at $z=0$. This mass dependence implies that the growth of 
SMBHs and galaxies in low-mass halos is almost frozen after $t_{\rm q}$, 
while the growth in massive halos is not.
The net effect is a ``stretch'' of the $M_{\rm BH}$-$M_*$ relation
towards the high-mass end from an anchor point at the low-mass end, 
as shown by the black dots in the left panel of Fig.~\ref{fig:scaling-init}. 
As mergers tend to give a 1:1 relation between the involved masses, 
the $M_{\rm BH}$-$M_{*}$ and $M_{\rm BH}$-$M_{\rm h}$ 
relations at the most massive ends steepen relative to those at 
$t_{\rm q}$. The history of the scaling relations is then erased
by mergers,  leaving behind little trace of the seeding and 
early growth of the SMBHs.

The TNG subgrid model for AGN feedback is designed to reproduce the 
observed $M_{\rm BH}$-$M_{\rm *,bulge}$ relation of \citet{Kormendy2013},  
which is a power-law form with an index of $\gamma \approx 1$.
As shown by \citet[see their figure 5]{weinbergerSimulatingGalaxyFormation2017}, the 
$M_{\rm BH}$-$M_{\rm *,bulge}$ relation of the TNG model roughly 
follows the $\gamma = 1$ relation, and so does the $M_{\rm BH}$-$M_*$ relation.
Since current observational results still have significant uncertainties in 
estimating the SMBH and bulge masses, it is still unclear whether or not 
the observed $M_{\rm BH}$-$M_*$ (or $M_{\rm *,bulge}$) relation can be extended to 
lower SMBH masses, and what the power index really is.
In a more recent work, \citet{greeneIntermediateMass2020}
suggested a steeper relation, $M_{\rm BH} \sim M_{*}^{1.33}$ for early-type 
galaxies. \citet[][see their table 2]{grahamAppreciatingMergersUnderstanding2023} obtained  
$M_{\rm BH} \sim M_{*}^{1.69}$ (and $M_{\rm BH} \sim M_{\rm *, bulge}^{1.64}$) 
for pure elliptical galaxies. 
For the $M_{\rm BH}$-$M_{\rm h}$ relation, \citet{voitBlackHoleGrowth2024}
showed that the prediction from TNG is significantly flatter than  
observations. For the model to match the observations, 
they suggested modifications of the criteria for the triggering
of the kinetic-mode feedback and/or the feedback efficiency. These 
more recent observations in the nearby Universe can thus put constraints on the 
subgrid physics to be implemented in the next generation of simulations,
especially for the SMBH growth in the phase of quenched evolution.

Since mergers in the quenched phase erased the imprints of early evolution,
in particular for SMBHs in massive halos, a slice of the 
$M_{\rm BH}$-$M_*$ (or $M_{\rm BH}$-$M_{\rm h}$) flow lines 
in the nearby Universe may not be adequate to infer the full tracks 
of galaxies.
This can be seen by comparing our TNG-based conclusions to those 
inferred by \citet[][see their figures 8 and 9]{grahamAppreciatingMergersUnderstanding2023}
from the observed $M_{\rm BH}$-$M_{\rm *,bulge}$ relations for galaxies
with different morphological types. 
At $M_{\rm *,bulge}\gtrsim 10^{11} \Msun$, their analysis suggests that
mergers between elliptical galaxies (with bulge-to-total ratio $\approx 1$) tend to 
drive migrations in the $M_{\rm BH}$-$M_{\rm *,bulge}$ plane towards a 1:1 relation, 
consistent with our inferences from the TNG. However, our analysis suggests that, 
for galaxies with $M_* \lesssim 10^{10.5}\Msun$, the ex-situ growth in both $M_*$ 
and $M_{\rm BH}$ contributes little to the total mass, which is inconsistent with 
the merger-driven scenario for the formation of elliptical galaxies and 
the build-up of the $M_{\rm BH}$-$M_{\rm *,bulge}$ relation in the low-mass end. 
One possibility is that the dynamical hotness of these galaxies comes 
from the fast inflow of gas associated with the fast regime of halo assembly 
\citep[e.g.][]{zhaoGrowth2003a}, as suggested by
\citet[][see also \citealt{yuBornThisWay2023}]{moTwophaseModelGalaxy2024}.
The change of the power index of the $M_{\rm BH}$-$M_*$ relation 
at $t_{\rm q}$, as demonstrated in this paper, is then expected to occur if
the kinetic-mode feedback drives the quenching of galaxies.

One way to circumvent the erasure of the long-term evolutionary history by mergers 
is to find alternative observational indicators in the nearby Universe that can 
separate in-situ and ex-situ components of the stellar mass.
As mergers tend to create dynamically hot stellar orbits, the in-situ 
and ex-situ growth of stellar mass may be disentangled by decomposing a 
galaxy into different dynamical components.
Indeed, using IllustrisTNG and EAGLE simulations, \citet{zhuMassDynamicallyHot2022}
analyzed the consequence of galaxy mergers and found that the mass of the 
``hot inner stellar halo'' is a good indicator of the stellar mass of ex-situ origin.
With high-quality IFU observations, such a decomposition can be made by, e.g. 
the maps of stellar mass, kinematics, metallicity, and age
\citep{zhuDisentanglingFormationHistory2020}. Scaling relations can then be built
between the SMBH mass and the masses of different stellar components.

On the other hand, observations directly targeting SMBHs at high $z$, 
where they are less massive and more active, are crucial to cover the early 
transitions/edges in the evolutionary history of the SMBH growth and to avoid the 
ambiguity in the extrapolation of the local scaling relations. Using the deep 
imaging and spectroscopy capabilities 
of JWST, \citet{wangStrongHeII2024} found a galaxy with a spectrum 
compatible with that produced by a Pop-III star cluster. As theoretical
studies \citep[e.g.][]{stacy2016BuildingPopulationIII,wiseFormationMassiveBlack2019,
chonSupermassiveStarFormation2020,
latifBirthMassFunction2022,latifTurbulentColdFlows2022,regan2024MassiveBlackHole,
ventura2024SemianalyticModellingPop} 
have suggested that massive Pop-III stars are promising SMBH seeds of 
intermediate masses ($\sim 10^4 \Msun$),
direct observations of Pop-III stars (clusters) 
will be crucial to distinguish among different seeding scenarios 
\citep[see, e.g.][]{spinosoMultiflavourSMBHSeeding2023} and to 
probe the growth of SMBHs in Phase-1.

A recent breakthrough in the observations of SMBHs at high $z$ with JWST
is the discovery of an abundant population of ``little red dots''
\citep[e.g.][]{mattheeLittleRedDots2024}. 
Most of these sources have spatial extents of less than $1\,{\rm kpc}$, and 
their gas densities are expected to be sufficiently high to support fast   
accretion by SMBHs. Indeed, the spectral features, 
as suggested by \citet{liLittleRedDots2024}, are consistent with low-mass 
SMBHs that accrete at about the Eddington rate and reside in extended, dusty environments 
produced by recent star formation and modified by the AGN feedback. 
These systems are thus likely to be SMBHs in their Phase-2
according to the subgrid model of TNG. Follow-up observations 
and analyses of these sources are needed to obtain the
SMBH mass function, accretion rates, and the properties of their host 
galaxies,  so as to determine when and how Phase-2 starts and 
ends.

\subsection{Inferring the paths of SMBH growth from observations} 
\label{ssec:obs-paths}

\begin{figure}[htb]
\centering
\includegraphics[width=.45\textwidth]{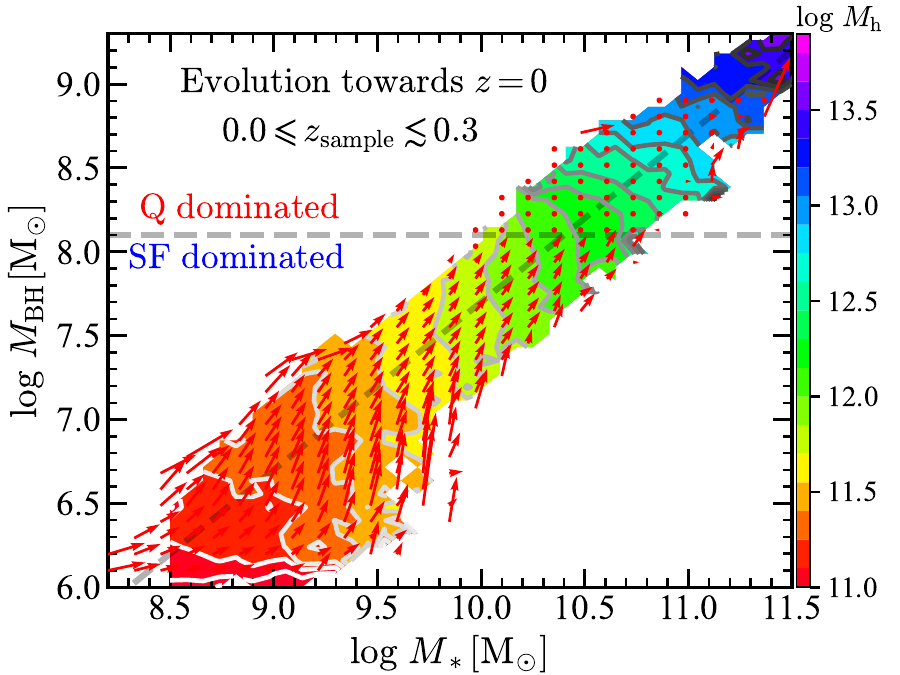}
\caption{
    Distribution and movement of galaxies in the $M_{\rm BH}$-$M_*$ plane.
    {\bf 2-D histogram} is shown for all central galaxies in TNG with 
    $M_* \geqslant 10^{8.5}\Msun$ at $z=0$. Colors are encoded and contours 
    are drawn to show their host halo masses.
    {\bf Red arrows} are path segments of these galaxies 
    from $z=0.3$ to $z=0$.
    Tilted {\bf dashed line} shows the $M_{\rm BH}$-$M_{\rm *,bulge}$ relation 
    found in observation by \citet{Kormendy2013}.
    See \S\ref{ssec:obs-paths} for a detailed discussion.
}
\label{fig:flow-plot-mbh-vs-ms} 
\end{figure}

The discussion in \S\ref{ssec:obs-scaling} only requires the measurements of 
masses of SMBHs, galaxies, and halos at different redshifts. The scaling 
relations built this way only focus on the time slices of the flow lines 
for a population of SMBHs, but miss the paths of individual SMBHs 
across these slices. One way to complete the missed information is to 
additionally incorporate the first derivatives of the masses: HAR, SFR and 
BHAR. Among them, both the SFR and BHAR can be directly measured, 
e.g. from the luminosities of the AGN and galaxy in some bands.
In contrast, HAR can only be indirectly inferred, e.g. from the galaxy size 
\citep{liangConnectionGalaxyMorphology2024}, if a well-calibrated empirical relation 
is available, or from constrained simulations, if the observed 
tracers of the underlying field are sufficiently fair and dense
\citep[e.g.][]{wangELUCIDEXPLORINGLOCAL2016,tullyCosmicflows3CosmographyLocal2019,boruahBayesianReconstructionDark2022,dolagSimulatingLOcalWeb2023,
sawalaDistinctDistributionsElliptical2024}, 
The inclusion of the derivatives allows a first-order extrapolation of 
the locations of SMBHs in the space of masses, and thus provides evolutionary 
path segments that carry dynamical information about SMBH growth.

Fig.~\ref{fig:flow-plot-mbh-vs-ms} shows the distribution of SMBHs in the
$M_{\rm BH}$-$M_*$ plane at $z=0$, 
with colors and contours encoded according to their host halo masses.
For each population of SMBHs with a given $M_{\rm BH}$ and $M_*$, an arrow is 
drawn to indicate the movement of its location in this plane since $z=0.3$. 
To cover larger ranges of $M_{\rm BH}$ and $M_*$ for this narrow 
redshift interval, we enlarge the sample size by including
all branches with $M_{*, z=0} \geqslant 10^{8.5} \Msun$.
Clear trends can be seen in different regions, each containing SMBHs in a 
given phase discussed in \S\ref{sec:processes}.
The lower left corner of the plane is populated with SMBHs in Phase-1, where 
$M_{\rm BH}$ is close to the seed mass of $\sim 10^6 \Msun$.
The arrows in the region of $M_* \sim 10^{9.5} \Msun$ and 
$M_{\rm BH} \lesssim 10^7 \Msun$ align with the vertical 
axis, which marks the transition from the star-formation dominated growth
to the SMBH-accretion dominated growth (Phase-2).
Once SMBHs approach the local scaling relation of $M_{\rm BH}$-$M_*$ 
from either the upper or lower side, the flow moves along this relation 
by the self-regulation dominated growth (Phase-3).
The prevalence of massive central SMBHs of $M_{\rm BH} \sim 10^{8.2} \Msun$ 
\citep[][see their figure 5]{terrazasRelationship2020} drives the quenching of 
galaxies and freezes the evolution of the SMBHs in the $M_{\rm BH}$-$M_*$ 
plane during Phase-4.  For the most massive SMBHs, mergers tend to drive a 
``dry'' evolution towards higher masses, 
although the sample size is not enough at the high-mass end to produce 
stable flow lines.

Drawing the flow lines in the $M_{\rm BH}$-$M_{\rm *}$ plane from observations
is more challenging, as it requires the measurements of four quantities 
$M_{\rm BH}$, $M_*$, BHAR and SFR. For a galaxy with inactive SMBH, BHAR 
is too low to be detected, while for a galaxy with active SMBH, a decomposition
of its image and spectrum is needed to separate the contributions from 
the AGN and stars. The available data thus only provide an incomplete 
view of the SMBH populations, together with measurement uncertainties
that probably mix flow lines. Thus, caution should be taken when 
interpreting the results in order to avoid biased conclusions.
A recent work by \citet[][see their figure 4]{zhuangEvolutionaryPathsActive2023}
indeed achieved this goal by stacking a large sample of AGNs at 
$z \leqslant 0.35$, with physical properties derived by a simultaneous 
decomposition of image and spectrum.
The set of flow lines they obtained converges to the local scaling relation, 
thus supporting the results produced by the subgrid model of TNG for the 
SMBH growth.
A noticeable difference between their results and those of TNG is the presence of 
over-massive black holes lying above the local scaling relation. 
This difference is either due to uncertainties in their measurements, 
or due to the strong feedback in TNG that forbids the growth of SMBHs 
beyond the local scaling relation. Future observations 
and more precise measurements are needed to pin down the exact reason.

An observationally more accessible way to infer the evolutionary paths
is to only use measurements of the mass of different components. One possibility is to
use the size of the scatter in the $M_{\rm BH}$-$M_{\rm *}$ 
plane to distinguish different phases.
This has been suggested by \citet{mcalpineRapidGrowthPhase2018}, who found that 
the EAGLE simulation produces a redshift-dependent critical halo mass for the 
SMBHs to transit from Phase-1 to Phase-2. As Phase-2 has rapid $M_{\rm BH}$
growth while Phase-1 and Phase-3 do not, the scatter in $M_{\rm BH}$ is expected to be large 
during Phase-2 relative to 
its two sides where the slopes of the $M_{\rm BH}$-$M_*$ 
and $M_{\rm BH}$-$M_{\rm h}$ relations are shallower.
Another possibility is to use multivariate statistical methods, such as 
the regression of $M_{\rm BH}$ on the joint of $M_*$ and $M_{\rm h}$.
This can be seen from the contours in 
Fig.~\ref{fig:flow-plot-mbh-vs-ms}: for $M_{\rm h} < 10^{12.3} \Msun$ (or $M_* < 10^{10.5}\Msun$), corresponding to  
$M_{\rm BH} < 10^8 \Msun$, the contours 
are almost vertical, {\revised indicating that $M_{\rm BH}$ does not depend 
on $M_{\rm h}$ once $M_*$ is fixed.}
At higher masses, the contours become horizontal, suggesting that 
the growth of the SMBH is related more to the halo than to the galaxy. Such a change 
is implied by the transition from Phase-3 (\S\ref{ssec:phase-3}), where 
the growth of SMBH is regulated by the thermal-mode feedback that balances the 
gravity of gas within the galaxy, to Phase-4 (\S\ref{ssec:phase-4}), 
where the growth of SMBH is reduced by the kinetic-mode feedback that balances 
the cooling of the whole central subhalo.
Thus, the scale on which the feedback energy is coupled to 
the surrounding gas is imprinted 
in the correlation between mass variables.
Indeed, \citet{zhang2024HaloMassobservableProxy} found an observational trend 
similar to the contours shown in Fig.~\ref{fig:flow-plot-mbh-vs-ms}, using $M_{\rm h}$ measured from galaxy-galaxy 
weak lensing and $M_{\rm BH}$ converted from stellar velocity dispersion. 
The transition mass scales they obtained, $M_{\rm h} 
\sim 10^{12.0} \Msun$ and $M_{\rm BH} \sim 10^{7.4} \Msun$, are consistent with our findings from TNG.

\section{Summary and Discussion}
\label{sec:summary-and-discussion}

\subsection{Main results}
\label{ssec:main-results}

In this paper, we used the IllustrisTNG simulation to quantify the 
co-evolution of SMBHs, their host galaxies, and halos. 
We focus on galaxies that are quenched at $z=0$ (\S\ref{sec:data-and-sample}) and use 
the main branches of the merger trees rooted in these galaxies to trace their evolution. 
We summarize our results and conclusions in the following.
\begin{enumerate}[topsep=0pt,parsep=0pt,itemsep=0pt,leftmargin=0pt,label=(\roman*)]
    
\item Although the growth histories of individual galaxies in the $M_{\rm BH}$-$M_*$ plane 
are diverse, they show regularities represented by four distinct phases. We fit 
the $M_{\rm BH}$-$M_*$ history of each galaxy in our sample with a function that allows us to 
separate the four phases and locate the transition points between them. 
We found that the first two phases are short, about $1$-$2\Gyr$ on average, while the 
last two phases are significantly longer. Each phase is found to be associated 
with a process that dominates the growth of the SMBH and its host galaxy,
and physical conditions can be identified for the transition from one phase to the next 
(\S\ref{sec:def-phases}).

\item Phase-1 is dominated by star formation and its feedback.
The starting point of the SMBH growth depends on the seeding strategy, and cannot be
fully constrained by observations currently available. 
Gas-rich environments resulted from the fast accretion of halos and effective 
cooling drive rapid star formation, but the SMBH growth is slow due to the 
low mass of the seed. The sSFR in this phase is higher than the sBHAR, and stellar feedback 
dominates the total feedback energy. The maximal duration of this phase is determined by the 
seed mass of SMBH and the gas environment (density and effective sound speed). 
A galaxy transits to Phase-2 when its sBHAR exceeds the sSFR (\S\ref{ssec:phase-1}).

\item Phase-2 is dominated by the accretion and thermal-mode feedback of 
the SMBH. The growth of the SMBH in this phase follows the exponential form 
expected from the Bondi accretion until the associated AGN feedback becomes sufficiently 
effective to unbind the gas of the galaxy. The effect of AGN feedback is inside-out,
strong on the surroundings of the SMBH and weaker on the whole galaxy and the halo.
Galaxies make transitions to Phase-3 when the difference between the cumulative 
energy gain from feedback and the energy loss due to gas cooling exceeds the binding energy 
of the gas in the galaxy (\S\ref{ssec:phase-2}).

\item Phase-3 is dominated by the self-regulation of SMBH growth.  
The galaxy in this phase is kept in a quasi-equilibrium state where the effective 
energy gain roughly balances the binding energy of the gas in the galaxy.
The regulation corresponds to power-law relations among the masses:
$\eta_{\rm eff} \eta_{\rm thermal} M_{\rm BH}
\sim M_{\rm g} V_{\rm g}^2  
\sim f_{\rm gas} M_{\rm h}^{5/3}
\sim f_{\rm gas} f_{\rm *}^{-5/3} M_*^{5/3}$. The exact values of 
the power-law indices depend on the gas fraction, the stellar mass fraction, 
and the feedback efficiency (\S\ref{ssec:smbh-self-regularization}).

\item 
The trigger of the kinetic-mode AGN feedback depends on halo mass.  
It is closely related to the transition of halo assembly from the fast to slow regimes  
for halos with $M_{{\rm h}, z=0} \lesssim 10^{12.5} \Msun$, and is much earlier
than the transition of halo assembly for more massive halos
(\S\ref{sssec:phase-3-triger-kin}).
The kinetic-mode AGN feedback is a necessary condition for the quenching of 
a central galaxy, but quenching can be delayed by mergers. 
A sufficient condition for the quenching of a galaxy is that the cooling
rate of the hot gas in the subhalo is comparable to the energy output rate of 
the AGN (\S\ref{sssec:quenching-criterion}).

\item Phase-4 is dominated by the ex-situ growth driven by galaxy mergers,
with in-situ growth playing a much smaller role. 
$M_*$ and $M_{\rm BH}$ of ex-situ origin assembled in this phase account
for more than $80\%$ of their final values for some galaxies.
The balance between the energy output rate of the AGN and the cooling rate within 
the subhalo is sustained during Phase-4 (\S\ref{ssec:phase-4}).

\item
The power index ($\gamma$) of the $M_{\rm BH}$-$M_*$ relation varies
as the dominating processes for the growth of SMBHs and galaxies change
between different phases. The value of $\gamma$ is about $1.5$ at $t_{23}$ due 
to the condition for the thermal-mode feedback to overcome the potential energy of the gas; 
it becomes smaller than $1$ at $t_{\rm q}$ due to the specific choice of TNG 
for triggering the kinetic-mode feedback; and it is driven back to $\approx 1$ 
at $z = 0$ for high-mass galaxies by mergers. We highlight potential observational 
tests for the predictions of the simulation. These include the use of scaling relations 
that bracket the phases and transitions, and the use of the mass growth rates to reconstruct 
the evolution paths of SMBHs and galaxies (\S\ref{sec:observation}).

\end{enumerate}

\subsection{Implications}
\label{ssec:implications}

We note that our results and conclusions are based on the numerical 
and subgrid models of IllustrisTNG. How the predicted observables respond 
to the change of model assumptions is of great interest and is 
one of the main targets in using hydrodynamical simulations to understand 
galaxy formation.
Such a question has been explored by \citet{habouzitSupermassiveBlackHoles2021}
using a set of different simulations: Illustris, IllustrisTNG, Horizon-AGN,
EAGLE and SIMBA. 
All these simulations produce a $M_{\rm BH}$-$M_*$ relation 
similar to that observed for high-mass ($M_* \gtrsim 10^{10.5}\Msun$)
galaxies at $z\sim 0$ \citep[e.g.][]{Kormendy2013}. Such an agreement is not
significant, however, as the observational relation was used to calibrate the model 
parameters in these simulations. More interesting and significant is 
that the paths of individual galaxies in the $M_{\rm BH}$-$M_*$ plane 
before they reach the local scaling relation are different both among 
different simulations and at different redshifts within a given simulation.

In EAGLE, the $M_{\rm BH}$-$M_*$ relation at both $z=3$ and $z=0$ is found 
to be non-linear (in logarithmic scale) and to drop at $M_*\lesssim 10^{10.5}\Msun$, 
reflecting the strong supernova feedback in suppressing the growth of SMBHs 
(corresponding to our Phase-1) in low-mass galaxies at both redshifts. 
As discussed in \S\ref{ssec:phase-1} and based on the results of 
\citet{mcalpineLinkGalaxyBlack2017}, the transition from Phase-1 to Phase-2
in EAGLE occurs when the virial temperature of the halo reaches a time-independent
critical value. The corresponding critical halo mass changes moderately 
from $M_{\rm h} \approx 10^{11.2}\Msun$ at $z=6$ to $10^{12.4}\Msun$
at $z=0$. At a given redshift, the observed $M_{\rm BH}$-$M_*$ relation
at the low-mass end is thus dominated by the insufficient growth of SMBHs in Phase-1,
which causes the relation to bend downwards with respect to the local scaling relation.

The situation in TNG is different in that the non-linearity of the 
$M_{\rm BH}$-$M_*$ relation at the low-mass end appears at $z=3$ but 
disappears at $z=0$. \citet{habouzitSupermassiveBlackHoles2021} 
attributed this to the ``inefficient'' SN feedback at low $z$ in TNG. 
We can follow the argument used in deriving $\tau_1$ in \S\ref{ssec:phase-1} to obtain some 
quantitative understanding, by explicitly incorporating the 
definition of the transition into the Bondi accretion. Equaling the sBHAR given by the 
Bondi formula (Eq.~\ref{eq:mbh-dot-phase-1}) to the sSFR, we obtain
the SMBH mass at the transition $t_{12}$ as
$M_{\rm BH}(t_{12}) = {\rm sSFR}(t_{12}) / A_{\rm s}(t_{12})$.
Because $A_{\rm s}$ can be approximated as a constant during Phase-1 
(see Fig.~\ref{fig:phase-1-mass-history}) and the sSFR of star-forming galaxies 
decreases by $\gtrsim 1\dex$ from $z=3$ to $z=0$ in TNG 
\citep[e.g.][see their figure 8]{chenMAHGICModelAdapter2021},
$M_{\rm BH}(t_{12})$ is expected to decrease by $\gtrsim 1\dex$
in this redshift range. As seen from Fig.~\ref{fig:fit-info},
this decrease is sufficient to push $M_{\rm BH}(t_{12})$ downwards  
to a position below the seed mass of SMBHs. Thus, a newly seeded SMBH at low $z$ 
skips the entire Phase-1, which makes the $M_{\rm BH}$-$M_*$ relation linear towards 
the low-mass end. Indeed, for galaxies at $z \leqslant 0.3$ 
shown in Fig.~\ref{fig:flow-plot-mbh-vs-ms}, the flow lines at $M_{\rm BH} \approx 10^6 \Msun$ 
go upwards without a significant horizontal movement, as expected from our argument. 
Our result also implies that, if a lower seed mass is adopted while keeping other recipes 
unchanged in TNG, or if the seeding is delayed until a higher halo/stellar mass is reached,  
the non-linear $M_{\rm BH}$-$M_*$ relation would appear at lower $z$. 
The latter condition is fulfilled in SIMBA, and a non-linear 
$M_{\rm BH}$-$M_*$ relation towards the low-mass end is indeed found 
at $z=0$ in this simulation
\citep[][see their figure 11]{habouzitSupermassiveBlackHoles2021}.

In more extreme cases where SN feedback is weak even at high $z$ or totally 
absent in the simulation, a newly seeded SMBH would immediately start its 
rapid growth and Phase-1 is expected to disappear completely. The weak SN 
feedback adopted by Horizon-AGN is such a case, and indeed the predicted 
$M_{\rm BH}$-$M_*$ relation at $z=3$ is roughly linear \citep[][see their figure 11]{habouzitSupermassiveBlackHoles2021}. 
Numerical experiments of disabling SN feedback by \citet[][see their figure 7]{bowerDark2017b} 
also produce such a linear relation.

The duration of Phase-2, $\tau_2$, as shown in Fig.~\ref{fig:phase-2-mass-history} 
and estimated by Eq.~\eqref{eq:mbh-dot-numeric}, is about $1\Gyr$. However, this 
timescale depends on the gas environment around a SMBH, and thus depends on the 
density threshold of star-forming gas assumed by the simulation, the subgrid models
evolving the cold gas, and the smoothing kernel defining the SMBH neighborhood. 
The threshold density is 
$n_{\rm gas} \approx 0.1 \perccm$ in both EAGLE and TNG, 
far below the density 
that a real galaxy can reach, as inferred both analytically and observationally
(see, e.g. \S2 of \citealt{chenTwophasePaper3-2024} and \S6 
of \citealt{dekelEfficientFormationMassive2023}). 
Lifting this threshold is computationally expensive and thus not feasible
for simulations on cosmological scales. Using a set of idealized simulations
for individual gas clouds embedded with randomly seeded low-mass BHs,
\citet{shiHyperEddingtonBlackHole2023} found that the 
small-scale structure of a gas cloud can be complex. The runaway collision of a 
BH with a gas sub-cloud produces a boost of $M_{\rm BH}$, 
much resembling the rapid growth of Phase-2. 
The collision can repeat multiple times, leading to a ladder-like
growth history of $M_{\rm BH}$ (see their figures 1 and 2).
Each collision starts and ends within a duration $\ll 1 \Myr$, more than 
three orders of magnitude shorter than our $\tau_2$, and even shorter 
than the lifetime of massive stars so that SNs are not able to 
be triggered.
On the other hand, \citet{hopkinsFORGEFIREII2024} conducted a 
``hyper-refinement'' zoom-in simulation for a galaxy within the cosmological 
context. Asymmetric gas structures around the SMBH are found and sustained 
super-Eddington accretion flow coexists with outflow. Star-formation within 
the influence radius of the SMBH is suppressed due to the pressure from 
turbulence and magnetic field, making star formation and stellar feedback ineffective.
The high density and complex sub-structure of gas,
the fast and repeated boost of $M_{\rm BH}$ by super- or hyper-Eddington 
accretion within a timescale less than $\Myr$, 
and the inefficient stellar feedback, all suggest rich physics of Phase-2 that needs to 
be explored further.

The ending of Phase-2 at $t_{23}$ by the AGN feedback to unbind 
the gas (Eq.~\ref{eq:scaling-t-23}) depends on the (radiative and coupling) 
efficiency of the feedback energy ($\eta_{\rm thermal}$) and 
subsequent cooling of the heated gas ($\eta_{\rm eff}$).
A simulation with different $\eta_{\rm thermal}$ and/or $\eta_{\rm eff}$
can thus produce a $M_{\rm BH}$-$M_*$ relation at $t_{23}$ 
(Fig.~\ref{fig:scaling-init}) with different normalizations.
Since the self-regulation of SMBHs during Phase-3 is also determined by 
these two factors, the normalization of the $M_{\rm BH}$-$M_*$ relation
during Phase-3 and at $t_{\rm q}$ will also be changed.
The difference in the normalization of the scaling relation shown in 
figure 4 of \citet{habouzitSupermassiveBlackHoles2021} for different
simulations can thus be understood along this line. On the other hand, the power index 
of the scaling relation at a specific event such as $t_{23}$ or $t_{\rm q}$, 
is distorted by the conditions for the event to be triggered. We 
showed in \S\ref{ssec:obs-scaling} that the kinetic-mode AGN 
feedback modeled in TNG is preferentially triggered in high-mass 
halos, which produces a shallow power index of the scaling relation 
at $t_{\rm q}$. Such a transition in SIMBA does not have a clear 
mass preference, and thus produces a more widely spread range of $M_*$ and 
a non-constant power index.

The final phase, Phase-4, is dominated by the ex-situ, gas-poor growth at 
the high-mass end. As the frequency of mergers is determined by the hierarchical 
structure formation in the $\Lambda$CDM cosmology, it is not sensitive 
to baryonic physics and is thus expected to be similar across simulations.
Indeed, all the simulations shown in figure 4 of \citet{habouzitSupermassiveBlackHoles2021} 
produce a tip at the high-mass end of the $M_{\rm BH}$-$M_*$ relation, with a 
power index of $\approx 1$. The normalization of this tip, however, inherits
the influences of baryonic physics at $t_{\rm q}$, and thus varies 
with simulations, as seen from the lower normalization of EAGLE. 
The scaling relation in Phase-4 for low-mass galaxies that have not experienced 
significant mergers is frozen after $t_{\rm q}$, and thus also depends on 
the baryonic physics.

In summary, our method based on separating the evolution of galaxies into 
different phases and identifying transitions between them 
provides a way to compare different simulations in a unified way. 
The identification of the driving processes in each phase provides a powerful 
avenue to connect observations to physical processes, and to improve the 
relevant subgrid models. We plan to come back to related investigations in the 
future.

\section{data availability}
Data directly related to this publication and its figures are available on
request from the corresponding author. The IllustrisTNG simulations themselves, 
are publicly available and accessible at \url{www.tng-project.org/data}.

\begin{acknowledgments}
We thank the referee for a useful report that significantly improved the paper.
This work is supported by the National Natural Science Foundation of China 
(NSFC, Nos. 12192224, and 11890693), CAS Project for Young Scientists 
in Basic Research (No. YSBR-062), the Fundamental Research Funds for the 
Central Universities, and the China Postdoctoral Science Foundation 
(No. 2022TQ0329).
We acknowledge the science research grants from the China Manned Space 
Project with No. CMS-CSST-2021-A03. 
The authors gratefully acknowledge the support of Cyrus Chun Ying Tang 
Foundations. 
HL thanks Kai Wang, Hui Hong, Xiong Luo, Yuan Wang, Enci Wang, Wentao Luo and Yu Rong for their valuable insights and discussions.
The work is also supported by the Supercomputer Center of University of 
Science and Technology of China. 
The authors would like to express their gratitude to 
Max Planck Computing and Data Facility (MPCDF) 
for providing the necessary computational and data storage resources that have 
significantly contributed to the research results presented in this paper.
\end{acknowledgments}

\appendix
\setcounter{figure}{0}
\renewcommand{\thefigure}{A\arabic{figure}}

\section{Fitting error}
\label{app:show_fit}

\begin{figure}[htb]
    \centering
    \includegraphics[width=.45\textwidth]{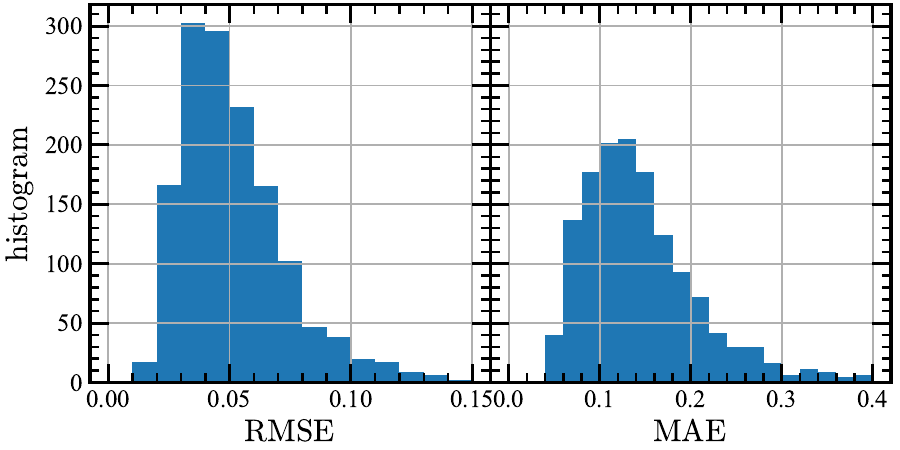}
    \caption{
        The distributions of root-mean-square error ({\bf left panel}) and 
        maximum absolute error ({\bf right panel}) of our fitting. 
    }
    \label{fig:fit-error} 
\end{figure}

The definition of the phases and transitions for each galaxy is based on the fitting 
of the $M_{\rm BH}$-$M_*$ relation in the evolutionary history (\S\ref{sec:def-phases}). 
Here we examine the quality of the fitting by two measurements of errors, the 
root-mean-square error (RMSE) and the maximum absolute error (MAE), defined as
\begin{align}
    {\rm RMSE}(y_{\rm sim},y_{\rm fit}) 
    &= \sqrt{\frac{1}{n} \sum_{s} \left(y_{{\rm sim},s}-y_{{\rm fit},s}\right) ^2 } 
    \,, \\
    {\rm MAE}(y_{\rm sim},y_{\rm fit}) 
    &= \underset{s}{\rm max} \left\{
        \left| y_{{\rm sim}, s}-y_{{\rm fit},s} \right|
    \right\}
    \,,
\end{align}
respectively,
where $y_{{\rm sim}, s}$ and $y_{{\rm fit}, s}$ are the simulated and best-fit 
values of $\log M_{\rm BH}$ at snapshot $s$.
The distributions of these two errors for our example of galaxies
(see \S\ref{sec:data-and-sample}) are shown in Fig.~\ref{fig:fit-error}.
The distribution of RMSE (MAE) peaks at $0.05$ ($0.1$), with most of the probability 
mass below $0.1$ ($0.3$). Given the large dynamical ranges of $M_{\rm BH}$ and 
$M_*$, the small errors suggest that our fitting safely captures the main 
trend of the evolutionary histories of individual galaxies.

\clearpage

\bibliography{ref}{}
\bibliographystyle{aasjournal}

\end{document}